\documentclass[a4paper,10pt, draftcls, onecolumn]{IEEEtran}
\usepackage{amssymb,amsmath}
\usepackage{amsmath}
\usepackage{threeparttable}   
\usepackage[dvips]{color}
\usepackage{url}   
\usepackage[T1]{fontenc}
\usepackage{algorithm}               
\usepackage{algorithmic}    
\usepackage{cite}
\usepackage{multirow}
\usepackage[dvips]{color}
\usepackage{lineno}
\usepackage[normalem]{ulem}
\usepackage{color,soul}  
\usepackage{indentfirst}
\usepackage{setspace}
\usepackage{cases}
\newcommand{\bm}[1]{\mbox{\boldmath{$#1$}}}

\usepackage{tikz}

\begin{document}
\title{Antenna Selection for Improving Energy Efficiency in XL-MIMO Systems}
\author{Jos\'e Carlos Marinello, Taufik Abr\~ao, Abolfazl Amiri, Elisabeth de Carvalho, \\  Petar Popovski 
\thanks{Copyright (c) 2015 IEEE. Personal use of this material is permitted.} 
\thanks{This work was supported in part by the  Arrangement between the European Commission (ERC) and the Brazilian National Council of State Funding Agencies (CONFAP), CONFAP-ERC Agreement H2020, by the National Council for Scientific and Technological Development (CNPq) of Brazil under grants 404079/2016-4 and 310681/2019-7.}
\thanks{J. C. Marinello is with Electrical Engineering Department, Federal University of Technology PR, Corn\'{e}lio Proc\'{o}pio, PR, Brazil;  \texttt{jcmarinello@utfpr.edu.br}.}
\thanks{T. Abr\~{a}o is with Electrical Engineering Department, State University of Londrina (UEL), Londrina, PR, Brazil;  \texttt{taufik@uel.br}.}
\thanks{A.  Amiri, E. de Carvalho and P. Popovski are with the Department of Electronic Systems,  Technical Faculty of IT and Design; Aalborg University,	Denmark; \texttt{petarp@es.aau.dk}.} 
}

\maketitle 

\begin{abstract}
We consider {the recently proposed} extra-large {scale massive} multiple-input multiple-output (XL-MIMO) {systems,} with some hundreds of antennas  serving a smaller number of users. Since the array length is of the same order {as} the distance to the users, the long-term fading coefficients of a given user vary with the different antennas at the {base station (BS)}. Thus, the signal transmitted by some antennas {might} reach the user with much more power than that transmitted by some others. From a green perspective, it is not effective {to simultaneously activate hundreds or even thousands of} antennas, since the power-hungry radio frequency (RF) chains of the active antennas increase significantly the total energy consumption. Besides, a larger number of selected antennas increases the power required by linear processing, such as precoding matrix computation, and short-term channel estimation. In this paper, we propose four antenna selection (AS) approaches to be deployed in XL-MIMO systems aiming at maximizing the total energy efficiency (EE). Besides, employing some simplifying assumptions, we derive a closed-form analytical expression for the EE of the XL-MIMO system, and propose a straightforward iterative method to determine the optimal number of selected antennas able to maximize it. The proposed AS schemes are based solely {on} long-term fading parameters, thus, the selected antennas set remains valid for a relatively large {time/frequency intervals}. Comparing the results, we find that the genetic-algorithm based AS scheme usually achieves the best EE performance, although our proposed highest normalized received power AS scheme also achieves very promising EE performance in a simple and straightforward way.
\end{abstract}



\section{Introduction}\label{sec:intro}
In {the fifth-generation (5G)} networks, massive multiple-input multiple-output (MIMO) {is} identified as a key technology for achieving large gains in spectral and energy efficiencies \cite{Marzetta16, Bjornson.2017book}. Recently, a new type of {very} large antenna arrays, which can be integrated into large structures like stadiums, or shopping malls, has been conceived: the so called extra-large scale massive MIMO (XL-MIMO) \cite{amiri2020deep, yang2019uplink, Heath19}. {XL-MIMO system is a very promising and recent technology, pointed out as important candidate for sixth-generation (6G) and beyond technologies \cite{Chen20a,Marzetta19a}, which is still in its inception, lacking for further elaborated techniques in order to mature the technology. Indeed,} due to the large dimension of the antenna array {in XL-MIMO systems}, different kinds of spatial non-stationarities appear accross the array {\cite{amiri2020deep, yang2019uplink, Heath19}}; hence, admitting constant long-term fading coefficients between a user and all the antennas of the array is not a valid assumption. {This is the main difference between the XL-MIMO scenario and the typical massive MIMO system model assumed in most part of massive MIMO literature. In \cite{Carvalho19}, it is shown through experimental measurements how different regions of an extremely large array see different propagation paths, and in some cases, the terminals might see just a portion of the array, called visibility region (VR). Authors also discuss how the non-stationarity properties of this new scenario change several important design aspects.}

In \cite{amiri2020deep} authors  seek for mapping users in terms of XL-MIMO array partition, such that the downlink (DL) sum-rate using a truncated zero-forcing (ZF) precoder is maximized.  Numerical results show that a properly trained network via deep learning approach solves the problem nearly as well as an optimal mapping algorithm. Hence, increasing the size of current massive MIMO arrays is promising in terms of boosting the spectral efficiency {(SE)} of the wireless systems. 
 
Since the centralized processing may present very high computational complexity in XL-MIMO arrays, a useful approach is to {split the signal} processing between subarrays. A subarray-based system architecture for XL-MIMO systems is proposed in \cite{yang2019uplink}, where closed-form uplink (UL) {SE} approximations with linear receivers are derived; the goal is to maximize the system sum achievable SE. Two statistical channel state information (CSI) based greedy user scheduling algorithms are developed, {providing} improved performance for XL-MIMO systems.

 In \cite{Heath19}, a simple {\it non-stationary channel model} is proposed for XL-MIMO systems, and the performance of conjugate beamforming (CB) and ZF in the DL have been investigated considering such channel. The non-stationarities are modeled in a binary fashion, such that each antenna can be visible or not for a {specific} user, giving rise to the VRs: {an area of the {massive antenna} array concentrating the {most of the received} user's energy}. However, the authors did not consider long-term fading variations between the visible antennas of a given user. 
 
{In \cite{amiri2020distributed} authors develop procedures for XL-MIMO receivers design. There are two important challenges in designing receivers for XL-MIMO systems: increased computational cost of the multi-antenna processing, and how to deal with the variations of user energy distribution over the antenna elements due to the spatial non-stationarities across huge distributed antenna-elements in the 2D or 3D array. Indeed, non-stationarities limit the XL-MIMO system performance. Hence, the authors propose a distributed receiver based on variational message passing that can address both challenges. In the proposed receiver structures,  the processing is distributed into local processing units, that can perform most of the complex processing in parallel, before sharing their outcome with a central processing unit. Such designs are specifically tailored to exploit the spatial non-stationarities and require lower computations than linear ZF or minimum mean square error (MMSE) receivers.}

{In \cite{Rodrigues20}, the ZF and regularized ZF schemes operating in XL-MIMO scenarios with a fixed number of subarrays have been
emulated using the randomized Kaczmarz algorithm (rKA), deploying non-stationary properties through VRs. Numerical results have shown that, in general, the proposed rKA-based combiner applicable to XL-MIMO
systems can considerably decrease computational complexity of the signal detector at the expense of small performance losses. On the other hand, in \cite{Wang20}, an expectation propagation detector for XL-MIMO systems has been proposed. In order to reduce complexity, the subarray-based architecture employed distributes baseband data from disjoint subsets of antennas into parallel
processing procedures coordinated by a central processing unit. Additionally, authors also propose strategies for further reducing the complexity and overhead of the information exchange between parallel subarrays and the central processing unit to facilitate the practical implementation of the proposed detector.}
 
Recently, to deal with subarrays and {channel} scatterers in non-stationary XL-MIMO environment,  \cite{Han_2020} proposed two {\it channel estimation methods} based on  subarray-wise and scatterer-wise  near-field non-stationary channel properties. Authors model the multipath channel with the last-hop scatterers under a spherical wavefront and divide the large aperture array into multiple subarrays.  
The proposed channel estimation methods {position} the scatterers and {perform a} mapping between subarrays and scatterers. Hence, the scatterer-wise method simultaneously positions each scatterer and detects its VR to further enhance the positioning accuracy. Moreover, the subarray-wise method can achieve low mean square error (MSE) performance under low-complexity, whereas the scatterer-wise method can accurately arrange the scatterers and determine the non-stationary channel.

In \cite{Tufvesson20a}, authors propose and validate realistic channel models when employing physically-large arrays, in which non-stationarities and visibility regions are present, as in the XL-MIMO system. The statistical distribution of important channel parameters are found based on measurements. Such contributions are proposed as extensions to the COST 2100 channel model. Besides, key statistical properties of the proposed extensions, e.g., autocorrelation functions, maximum likelihood estimators, and Cramer-Rao bounds, are derived and analyzed. Furthermore, the performance of a spatial modulation massive MIMO system is investigated in \cite{Fu20} under a non-stationary channel model. Authors show that spatial modulation can outperform typical employed spatial multiplexing transmission in certain scenarios of low correlation among sub-channels, for example under a rich scattering environment.

{A novel random access (RA) protocol for crowded XL-MIMO systems is proposed in \cite{Nishimura20}. Authors have proposed a decentralized and uncoordinated decision rule, which can be evaluated at the users side, for retransmitting or not the RA pilots during the connection stage, taking advantage of the XL-MIMO propagation features. The proposed protocol achieves significant performance improvements in terms of reducing the connection delay and providing access for larger number of devices.}

\subsection{Motivation, Contributions and Novelties in Comparison with Existing Works}

Current design approaches in telecommunication systems include a global effort in saving energy and reducing pollution \cite{Bjornson.2017book}, \cite{Marinello19}, \cite{Debbah15}. We show in this paper that antenna selection (AS) methods in XL-MIMO systems is a very important issue since the energy expenditure of such systems could be very high if activating the radio frequency (RF) chains of all antennas simultaneously. Besides, some antennas might contribute very little with the system performance due to the non-stationarities and visibility regions, in such a way that the power required to activate their RF chain becomes a burden that severely penalizes the total energy efficiency (EE) of the system. Therefore, the very large number of antennas deployed in the XL-MIMO systems in conjunction with the spatial non-stationarities make the application of AS schemes very important.

 The main \emph{contributions} of this work are threefold:
 \begin{itemize}
     \item[(\textbf{i})] {Reformulating the signal to interference plus noise ratio (SINR) performance expressions of \cite{Heath19}, considering long-term fading variations across the array and incorporating the maximum transmit power constraint into the expressions for CB and ZF, and finding more compact and comprehensive results, readily applicable for antenna selection procedures.}
     \item[(\textbf{ii})] {Based on the obtained expressions, and on a realistic power consumption model, we evaluate the total EE of the XL-MIMO system. Besides, we propose and compare four low-complexity AS procedures aiming to maximize the total EE of the system, different than \cite{amiri2020deep, yang2019uplink} which proposed SE-based AS schemes. Our proposed schemes are based solely on the long-term fading parameters, and the obtained solutions remain valid for larger time/frequency intervals.}
     \item[(\textbf{iii})] {Based on our proposed AS schemes, and some simplifying assumptions, we {derive} approximated closed-form EE expressions, and propose an iterative method for finding the optimal number of selected antennas which maximizes EE. Finally, numerical simulations have validated the proposed performance expressions and compared the different XL-MIMO AS schemes.}
 \end{itemize}
 
{AS methods for typical spatially stationary massive MIMO systems \cite{Alouini17, Sun17} is a well investigated topic. However, the XL-MIMO system is a different scenario. While the spatially stationary model applies for typical cellular systems, where the BS antenna array dimension is much lower than the distance to the users and a single long-term fading coefficient holds for all antennas, significant power variations appear along the XL-MIMO array, due to its large dimension and number of antennas, and proximity with users. The non-stationary XL-MIMO scenario just very recently was introduced in the literature. To the best of our knowledge, this contribution is the first evaluating the EE of the XL-MIMO scenario, showing that AS methods are especially important to improve EE due to the spatial non-stationarities that naturally arise in XL-MIMO systems, proposing long-term fading based AS procedures, and deriving the optimal number of active antennas for this new wireless communication context.}

{With respect to the existing XL-MIMO literature, we can point out as the \emph{main novelties} of our paper: although our system model and CB and ZF performance expressions are similar to that of \cite{Heath19}, authors have considered a binary visibility region model for the XL-MIMO scenario, in which no long-term fading  variation  occurs  for  the  visible  antennas.  Besides,  performance  expressions  are  dependent  of  power  coefficients obtained resolving a separated optimization problem for meeting power constraint, and no antenna selection is considered. Differently, we incorporated the power constraint into the performance expressions, arriving at more compact and comprehensive results, readily applicable for AS procedures, and considered long-term fading variations along the array. Besides, AS for XL-MIMO systems has been investigated only in \cite{amiri2020deep, yang2019uplink} at the moment of writing this paper; however, both works proposed SE-based AS schemes for XL-MIMO systems. Differently, based on only long-term fading coefficients, we propose AS schemes aiming to maximize the XL-MIMO total EE, since this is a very important issue due to the very large number of antennas at the XL-MIMO array, and the non-stationarities and visibility regions which arise in this scenario. Furthermore, the long-term fading approach has the advantages of being simpler than short-term ones, and of providing solutions which remain valid for larger time periods and all subcarriers (if employing a wideband system), reducing the computational complexity of the antenna selection approach and simplifying hardware due to switching and RF chain on-off requirements.} 

 \noindent{\textit{Notations:} Boldface lower and upper case symbols represent vectors and matrices, respectively. ${\bf I}_N$ denotes the identity matrix of size $N$, while $\{\cdot\}^T$ and $\{\cdot\}^H$ denote the transpose and the Hermitian transpose operator, respectively. We use $\mathcal{CN}(m, \sigma^2)$ {when referring} to a circular symmetric complex Gaussian distribution with mean $m$ and variance matrix $\sigma^2$. Besides, ${\rm tr}(\cdot)$ and ${\rm diag}(\cdot)$ are the trace and diagonal matrix operators, respectively, while $[{\bf A}]_{i,j}$ holds to the element in the $i$th row and $j$th column of matrix ${\bf A}$, and ${\bf a}_i$ refers to its $i$th column vector}.
 
\section{System Model}\label{sec:syst}
We consider a base station (BS) equipped with a linear XL-MIMO array with $M$ antennas uniformly distributed along a length of $L$ meters{, Fig. \ref{fig:SystModel}. In front of the extra-large array structure}, $K$ users are randomly distributed in a rectangular area, of length $L$ in the array parallel dimension, and with a distance to the array in the range $[0.1 \cdot L, L]$\footnote{{In order to guarantee a minimum distance of the users to the XL-MIMO array, as in \cite{Marinello19, Rodrigues20}.}}. Since the distances of the users to the antennas is of the same order of the array length $L$ the average received power varies along the XL-MIMO array, and therefore we cannot consider a single long-term fading coefficient for a given user {\cite{amiri2020deep, Carvalho19}}. Instead, we consider a long-term fading coefficient $\beta_{m,k}$ regarding the $m$-th antenna of the XL-MIMO array and the $k$-th user, {similarly as in \cite{amiri2020deep, amiri2020distributed, Rodrigues20, Nishimura20}}, given by
\begin{equation}\label{eq:beta}
\beta_{m,k} = q \cdot d_{m,k}^{-\kappa},
\end{equation}
in which $q$ is a constant determining the path loss in a reference distance, $d_{m,k}$ is the distance between the $m$-th antenna of the XL-MIMO array and the $k$-th user, and $\kappa$ is the path loss decay exponent. The channel matrix ${\bf H} \in \mathbb{C}^{M\times K}$ is thus formed by elements $h_{m,k} = \sqrt{\beta_{m,k}}\cdot \underline{h}_{m,k}$, in which $\underline{h}_{m,k} \sim \mathcal{CN}(0,1)$, {assuming a rich scattering environment as in {\cite{yang2019uplink, Heath19}}. If we arrange the long-term fading coefficients of a user in a diagonal matrix:
\begin{equation}\label{eq:Rk}
{\bf R}_k = {\rm diag}([\beta_{1,k}, \beta_{2,k},\ldots,\beta_{M,k}]) \in \mathbb{R}^{M \times M},
\end{equation}
and the elements $\underline{h}_{m,k}$ in a vector ${\bf \underline{h}}_{k} \in \mathbb{C}^{M \times 1}$, we have that each column of ${\bf H}$ can be defined as ${\bf h}_{k} = {\bf R}_k^{\frac{1}{2}} {\bf \underline{h}}_{k}$ {as in \cite{Heath19}}.

\begin{figure}[!htbp]
\centering
\includegraphics[width=0.4\textwidth]{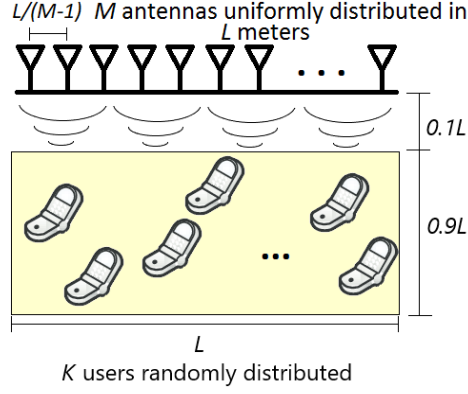} 
\vspace{-2mm}
\caption{Illustration of the adopted system model.}%
\label{fig:SystModel}
\end{figure} 

In the DL, considering an average received signal-to-noise ratio (SNR) $\rho$ at the users, an average long-term fading coefficient $\beta_{\rm avg}$ (among all antennas and users' positions), and a uniform power allocation policy for the users, the total transmit power, $P_{\rm max}$, should satisfy {\cite{Marzetta16}}
\begin{equation}
\rho = \frac{P_{\rm max} \cdot \beta_{\rm avg}}{\sigma^2},
\end{equation}
in which $\sigma^2$ is the noise power. Since the channel gain $\beta_{m,k}$ varies significantly along the array, it is more effective to select just the stronger antennas to transmit signal to the $k$-th user, reducing the number of active antennas{, as well} the power spent with power-hungry RF chains. We discuss in the next Section different approaches to obtain the set of antennas selected to serve the users, $\mathcal{A}$. For simplicity, we considered $\beta_{\rm avg} {\approx} q \cdot L^{-\kappa}$ in our simulations. 
The signal for user $k$, $s_k$, is precoded by ${\bf g}_k \in \mathbb{C}^{M \times 1}$ and scaled by $p_k \geq 0$, which adjusts the signal power, before transmission. {Considering a similar XL-MIMO system model than \cite{Heath19}, the} transmit vector $\bf x$ is the linear combination of the precoded and scaled signal of all the users, \emph{i.e.},
\begin{equation}\label{eq:transmit_sig}
{\bf x} = \sum_{k=1}^{K} \sqrt{p_k} \cdot {\bf g}_{k} \cdot s_k.
\end{equation}

Let ${\bf G} = [{\bf g}_1,{\bf g}_2,\ldots,{\bf g}_K] \in \mathbb{C}^{M \times K}$ be the combined precoding matrix, and ${\bf P} = {\rm diag}([p_1, p_2,\ldots,p_K]) \in \mathbb{R}^{K \times K}$ be the diagonal matrix of signal powers. The combined precoding matrix ${\bf G}$ is normalized to satisfy the power constraint
\begin{equation}\label{eq:powconst}
    \mathbb{E}[||{\bf x}||^2] = {\rm tr}({\bf P} {\bf G}^H {\bf G}) = P_{\rm max}.
\end{equation}

The signal received by the $k$-th user is
\begin{equation}
y_{k} = {\bf h}_{k}^H {\bf x} + n_{k}, \quad k = 1,2,\ldots K,
\end{equation}
in which $n_{k} \sim \mathcal{CN}(0,\sigma^2)$ is an additive white Gaussian noise (AWGN) sample. Assuming independent Gaussian signaling, \emph{i.e.}, $s_{k} \sim \mathcal{CN}(0,1)$ and $\mathbb{E}[s_i s_j^*] = 0$, $i \neq j$, the SINR $\gamma_k$ of the $k$-th user can be defined as {\cite{Heath19}}:
\begin{equation}\label{eq:SINR}
\gamma_k = \frac{p_k |{\bf h}_k^H {\bf g}_k|^2}{\sum_{j=1, j\neq k}^{K} p_j |{\bf h}_k^H {\bf g}_j|^2 + \sigma^2}. 
\end{equation}

We selected the CB and  ZF approaches as representative low-complexity linear precoding schemes. The CB precoder matrix is simply defined as
\begin{equation}\label{eq:CBprec}
    {\bf G}_{\textsc{CB}} = \alpha_{\textsc{CB}} {\bf H},
\end{equation}
and the ZF precoding matrix is
\begin{equation}\label{eq:ZFprec}
    {\bf G}_{\textsc{ZF}} = \alpha_{\textsc{ZF}} {\bf H} ({\bf H}^H {\bf H})^{-1},
\end{equation}
where the scaling factors $\alpha_{\textsc{CB}} = \sqrt{P_{\rm max}/{\rm tr}({\bf P}{\bf H}^H {\bf H})}$ and $\alpha_{\textsc{ZF}} = \sqrt{P_{\rm max}/{\rm tr}({\bf P}({\bf H}^H {\bf H})^{-1})}$ ensure that the power constraint \eqref{eq:powconst} is met.

Using \eqref{eq:CBprec} in \eqref{eq:SINR}, the SINR of the $k$th user for CB is
\begin{equation}\label{eq:SINR_CB}
\gamma_k^{(\textsc{CB})} = \frac{p_k |{\bf h}_k^H {\bf h}_k|^2}{\sum_{j=1, j\neq k}^{K} p_j |{\bf h}_k^H {\bf h}_j|^2 + \frac{\sigma^2}{P_{\rm max}} {\rm tr}({\bf P}{\bf H}^H {\bf H})}. 
\end{equation}
Similarly, using \eqref{eq:ZFprec} in \eqref{eq:SINR}, the SINR of the $k$th user for ZF is
\begin{equation}\label{eq:SINR_ZF}
\gamma_k^{(\textsc{ZF})} = \frac{p_k P_{\rm max}}{\sigma^2 {\rm tr}({\bf P}({\bf H}^H {\bf H})^{-1})}. 
\end{equation}

Given the system model presented {in this Section in eq. \eqref{eq:beta}--\eqref{eq:SINR_ZF}}, and the deterministic equivalent analysis of \cite{Debbah13}, it is presented in \cite{Heath19} the deterministic equivalent of $\gamma_k^{(\textsc{CB})}$ in \eqref{eq:SINR_CB} as
\begin{equation}\label{eq:SINR_CB_de}
\overline{\gamma}_k^{(\textsc{CB})} = \frac{p_k ({\rm tr} ({\bf R}_k))^2}{\sum_{j=1, j\neq k}^{K} p_j {\rm tr} ({\bf R}_k {\bf R}_j) + \frac{\sigma^2}{P_{\rm max}} \sum_{j=1}^{K} p_j {\rm tr}({\bf R}_j)}, 
\end{equation}
and the deterministic equivalent of $\gamma_k^{(\textsc{ZF})}$ in \eqref{eq:SINR_ZF} as
\begin{equation}\label{eq:SINR_ZF_de}
\overline{\gamma}_k^{(\textsc{ZF})} = \frac{p_k P_{\rm max}}{\sigma^2 \sum_{i=1}^{K} p_i \left( {\rm tr} ({\bf R}_i) - \sum_{j=1, j\neq i}^{K} \frac{{\rm tr} ({\bf R}_i {\bf R}_j)}{{\rm tr} ({\bf R}_j)} \right)^{-1}}.
\end{equation}
where ${\bf R}_i$ is defined as in \eqref{eq:Rk}.

Having found the SINR of the $k$th user, the spectral efficiency is readily obtained as $\eta^s_k = \log_2(1+\gamma_k)$. On the other hand, the energy efficiency is {\cite{Marinello19, Debbah15}}
\begin{equation}\label{eq:eneff}
\eta_e = \frac{B \sum_{k=1}^K \eta^s_k}{\mathcal{P}},
\end{equation} 
in which $B$ is the system bandwidth, and $\mathcal{P}$ is the total power consumption, discussed in Section \ref{sec:power}.

\subsection{Further Advances {in} the Performance Expressions}\label{sec:advances}
We revisit the performance expressions for non-stationary XL-MIMO discussed in \cite{Heath19}, while propose further elaborations to arrive at lean and more comprehensive results. Note that the results of \eqref{eq:SINR_CB_de} and \eqref{eq:SINR_ZF_de} {depend} on the signal powers in both numerator and denominators, and such coefficients should be chosen in order to satisfy the power constraint in \eqref{eq:powconst}. In the simulation code made available by the authors of \cite{Heath19}, they apply the CVX solver of \cite{cvx} to find a matrix $\bf P$ satisfying \eqref{eq:powconst}. This makes the performance expressions less intuitive, while limiting the application of {AS} schemes as proposed in Section \ref{sec:contrib} of this paper. Hence, {in this subsection,} we shed light on deriving self-contained closed-form SINR expressions recalling the channel hardening massive MIMO properties. For that, we first rewrite \eqref{eq:powconst} in the following form:
\begin{equation}\label{eq:powconst2}
    \mathbb{E}[||{\bf x}||^2] = {\rm tr}({\bf P} {\bf G}^H {\bf G}) = \sum_{k=1}^{K} p_k ||{\bf g}_k||^2 = P_{\rm max}.
\end{equation}

If a uniform power allocation scheme is applied, the following equality holds
\begin{equation}\label{eq:powconst3}
    p_k ||{\bf g}_k||^2 = \frac{P_{\rm max}}{K}, \quad k=1,2,\ldots K.
\end{equation}
{Hence,} when adopting CB, {eq.} \eqref{eq:powconst3} becomes
\begin{equation}\label{eq:powconst3_cb}
    p_k \alpha_{\textsc{CB}}^2 ||{\bf h}_k||^2 = \frac{P_{\rm max}}{K}, \quad k=1,2,\ldots K,
\end{equation}
and we have an undetermined system with $K$ equations and $K+1$ variables. By choosing $\alpha_{\textsc{CB}} = 1$ for simplicity, the $p_k$ coefficients can be obtained for CB as
\begin{equation}\label{eq:pk_cb}
    p_k^{(\textsc{CB})} = \frac{P_{\rm max}}{K ||{\bf h}_k||^2}, \quad k=1,2,\ldots K.
\end{equation}

Following similar assumptions as in \cite{Heath19}, we have that
\begin{equation}
    ||{\bf h}_k||^2 = {\bf h}_k^H {\bf h}_k = {\bf \underline{h}}_k^H {\bf R}_k {\bf \underline{h}}_k \xrightarrow{M \to \infty} {\rm tr}({\bf R}_k),
\end{equation}

and a deterministic equivalent of \eqref{eq:pk_cb} is
\begin{equation}\label{eq:pk_cb_de}
    p_k^{(\textsc{CB})} = \frac{P_{\rm max}}{K {\rm tr}({\bf R}_k)}, \quad k=1,2,\ldots K.
\end{equation}

Substituting \eqref{eq:pk_cb_de} in \eqref{eq:SINR_CB_de}, we arrive at
\begin{equation}\label{eq:SINR_CB_de2}
\overline{\gamma}_k^{(\textsc{CB})} = \frac{{\rm tr} ({\bf R}_k)}{\sum_{j=1, j\neq k}^{K} \frac{{\rm tr} ({\bf R}_k {\bf R}_j)}{{\rm tr} ({\bf R}_j)} + \frac{K \sigma^2}{P_{\rm max}}}. 
\end{equation}

On the other hand, {for the case of} ZF, \eqref{eq:powconst} becomes
\begin{eqnarray}\label{eq:powconst2zf}
    \mathbb{E}[||{\bf x}||^2] &=& {\rm tr}\left({\bf P} {\bf G}^H {\bf G}\right) = P_{\rm max},\nonumber\\ &=&\alpha_{\textsc{ZF}}^2 {\rm tr}\left({\bf P} ({\bf H}^H {\bf H})^{-1}  {\bf H}^H {\bf H} ({\bf H}^H {\bf H})^{-1}\right) = P_{\rm max},\nonumber\\
    &=&\alpha_{\textsc{ZF}}^2 {\rm tr}\left({\bf P} ({\bf H}^H {\bf H})^{-1}\right) = P_{\rm max},\nonumber\\
    &=&\alpha_{\textsc{ZF}}^2 {\rm tr}\left({\bf P} {\bf V} \right) = P_{\rm max},
\end{eqnarray}
in which the matrix ${\bf V}$ is a diagonal matrix formed by the main diagonal elements of $({\bf H}^H {\bf H})^{-1}$. We can thus rewrite \eqref{eq:powconst2zf} as
\begin{equation}\label{eq:powconst2zf2}
    \alpha_{\textsc{ZF}}^2 \sum_{k=1}^{K} p_k [{\bf V}]_{k,k} = P_{\rm max},
\end{equation}
and if a uniform power allocation is employed
\begin{equation}\label{eq:powconst3a}
    \alpha_{\textsc{ZF}}^2 \, p_k \, [{\bf V}]_{k,k} = \frac{P_{\rm max}}{K}, \quad k=1,2,\ldots K.
\end{equation}

Again, making $\alpha_{\textsc{ZF}} = 1$, the $p_k$ coefficients can be obtained for {the ZF precoding} as
\begin{equation}\label{eq:pk_zf}
    p_k^{(\textsc{ZF})} = \frac{P_{\rm max}}{K \, [{\bf V}]_{k,k}}, \quad k=1,2,\ldots K.
\end{equation}
Following the analysis of \cite[App. A]{Heath19}, it can be shown that
\begin{equation}
    [{\bf V}]_{k,k}\xrightarrow{M \to \infty} \left({\rm tr}({\bf R}_k) - \sum_{j=1, j\neq k}^{K} \frac{{\rm tr} ({\bf R}_k {\bf R}_j)}{{\rm tr} ({\bf R}_j)} \right)^{-1},
\end{equation}
and a deterministic equivalent of \eqref{eq:pk_zf} is
\begin{equation}\label{eq:pk_zf_de}
p_k^{(\textsc{ZF})} = \frac{P_{\rm max}}{K} \left({\rm tr}({\bf R}_k) - \sum_{j=1, j\neq k}^{K} \frac{{\rm tr} ({\bf R}_k {\bf R}_j)}{{\rm tr} ({\bf R}_j)} \right), \,\, k=1,\ldots K.
\end{equation}

Substituting \eqref{eq:pk_zf_de} in \eqref{eq:SINR_ZF_de}, we arrive at
\begin{equation}\label{eq:SINR_ZF_de2}
\overline{\gamma}_k^{(\textsc{ZF})} = \frac{P_{\rm max}}{K \sigma^2} \left({\rm tr}({\bf R}_k) - \sum_{j=1, j\neq k}^{K} \frac{{\rm tr} ({\bf R}_k {\bf R}_j)}{{\rm tr} ({\bf R}_j)} \right). 
\end{equation}

{Equations \eqref{eq:SINR_CB_de2} and \eqref{eq:SINR_ZF_de2} show the XL-MIMO DL system performance employing CB and ZF, respectively, as further extensions of eq. \eqref{eq:SINR_CB_de} and \eqref{eq:SINR_ZF_de} from \cite{Heath19}. This is a first contribution of this manuscript, which serves as basis for the following EE and AS analysis.}

\noindent{\it Remark 1:} Although we have considered $\alpha_{\textsc{CB}} {= \alpha_{\textsc{ZF}}} = 1$ in our analysis, any other choice for these parameters would result in the same expressions, since would affect every numerator and denominator terms in the same {way}.

\noindent{\it Remark 2:} The SINR performance expressions presented in \cite[Table I]{Heath19} can be seen as particular cases of \eqref{eq:SINR_CB_de2} and \eqref{eq:SINR_ZF_de2} when neglecting long-term fading and applying the normalization ${\rm tr}({\bf R}_k) = {\rm tr}({\bf \Theta}_k) = M$ or ${\rm tr}({\bf \Theta}_k) = D$, where ${\bf \Theta}_k$ and $D$ are the matrix describing the VR of $k$th user and the number of visible antennas per user, respectively, as in \cite{Heath19}.

\subsection{Antenna Selection Model}\label{sec:asmodel}
Given our deterministic equivalent performance expressions for CB and ZF in eq. \eqref{eq:SINR_CB_de2} and \eqref{eq:SINR_ZF_de2}, respectively, we {can} rewrite these {expressions considering the activation subset} of antennas. {Hence, denoting $\mathcal{A}$ as the set containing the indices of the active antennas, the deterministic equivalent SINR for the CB precoding results}
\begin{equation}\label{eq:SINR_CB_as}
\overline{\gamma}_k^{(\textsc{CB})} = \frac{\sum_{m \in \mathcal{A}} \beta_{m,k}}{\sum_{j=1, j\neq k}^{K} \frac{\sum_{m \in \mathcal{A}} \beta_{m,k} \beta_{m,j}}{\sum_{m \in \mathcal{A}} \beta_{m,j}} + \frac{K \sigma^2}{P_{\rm max}}}, 
\end{equation}
while for the ZF:
\begin{equation}\label{eq:SINR_ZF_as}
\overline{\gamma}_k^{(\textsc{ZF})} = \frac{P_{\rm max}}{K \sigma^2} \left(\sum_{m \in \mathcal{A}} \beta_{m,k} - \sum_{j=1, j\neq k}^{K} \frac{\sum_{m \in \mathcal{A}} \beta_{m,k} \beta_{m,j}}{\sum_{m \in \mathcal{A}} \beta_{m,j}} \right). 
\end{equation}

{It is worth to note that, in our formulation, the {\it activation subset of antennas} is the same for all users, differently from \cite{amiri2020deep}, in which each user has its own set of active antennas aiming to maximize the system sum-rate. We justify our formulation since, when aiming to maximize the total energy efficiency, once the power-hungry RF chain of an antenna is active, it is better to take full advantage of it, transmitting signal for all users. It has no significant benefit in defining the activation subset of antennas in a per-user fashion, since the ZF approach {is able to} eliminate the inter-user interference, {while the power increment necessary} to compute the precoding vector with a slightly large number of antennas is small if compared to the power to activate the RF chain of the additional antenna, as {evinced} in the next subsection. Besides, it would result in more complicated performance expressions, probably in terms of short-term fading coefficients, and the dimension of the search space of the AS algorithms would scale with $K$, becoming considerably more complex and power consuming.}

\subsection{Power Consumption Model}\label{sec:power}

We follow the same power consumption model of \cite{Marinello19}, which is very similar to that in \cite{Debbah15}, and is a very realistic model. However, as we focus on the DL transmission, we do not consider the UL data rates as well as the UL transmit powers. In the XL-MIMO scenario analysed herein, we consider the power expenditures of the irradiated DL data signal (with the amplifier efficiency), $P_{\textsc{tx}}^{\textsc{dl}}$, the UL training, $P_{\textsc{tx}}^{{\text{tr}}}$, the channel estimation, $P_{\textsc{ce}}$, the coding/decoding, $P_{\textsc{c/d}}$, the backhaul, $P_{\textsc{bh}}$, the linear processing computation, $P_{\textsc{pr}}$, the transceiver chains, $P_{\textsc{tc}}$, and a fixed quantity regarding the circuitry power consumption required for site-cooling, control signaling, and load-independent power of backhaul infrastructure and baseband processors, $P_{\textsc{fix}}$. {Thus, the overall power consumption results}
\begin{equation}\label{eq:cp_model}
\mathcal{P} = P_{\textsc{tx}}^{\textsc{dl}} + P_{\textsc{tx}}^{{\text{tr}}} + P_{\textsc{ce}} + P_{\textsc{c/d}} + P_{\textsc{bh}} + P_{\textsc{pr}} + P_{\textsc{tc}} + P_{\textsc{fix}}.
\end{equation}

Our objective here is to investigate the dependence of the selected {subset of} antennas, $\mathcal{A}$, with the total energy efficiency of the system. Note that the total energy efficiency of the system depends on $\mathcal{A}$ in {different ways}. First, the sum rate of the system depends on the {SE} of the users, which is a function of their SINRs dependent of $\mathcal{A}$. {Moreover, the} sum rate impacts on the power expenditures of the coding/decoding, and the backhaul. Besides, the power consumption of the transceiver chains is modeled as
\begin{equation}
P_{\textsc{tc}} = P_{\textsc{syn}} + |\mathcal{A}| P_{\textsc{bs}} + K P_{\textsc{mt}},
\end{equation}
in which $P_{\textsc{syn}}$ is the power of the local {oscillator}, $P_{\textsc{bs}}$ is the power required to each active BS antenna operate, while $P_{\textsc{mt}}$ is the power required to each single-antenna mobile terminal (MT) operate. Note that $M$ is usually very high in an XL-MIMO system\footnote{Typically hundreds or even thousands of antennas.}, {while} $P_{\textsc{bs}}$ accounting for the power-hungry RF chains is considered in \cite{Marinello19} as 1 W per antenna. Thus, activating {the RF chains of all} BS antennas would result in a very large power expenditure, in such a way that it is very important to perform {a suitable} antenna selection procedure.

The power consumed with processing, $P_{\textsc{pr}}$, corresponds to the power required to obtain the transmit signal in \eqref{eq:transmit_sig}, to obtain the precoding matrix, and to obtain the {AS} set. Note that this power is also dependent on the number of active antennas $|\mathcal{A}|$. Following the model {in} \cite{Marinello19}, but including the {term of power related to} the {AS processing}, we have
\begin{equation}\label{eq:pproc}
P_{\textsc{pr}} = B \left( 1 - \frac{\tau}{\mathcal{S}} \right) \frac{\mathcal{C}_{\rm ts}}{\mathcal{L}_{\textsc{bs}}} + \frac{B}{\mathcal{S}} \frac{\mathcal{C}_{\rm prec}}{\mathcal{L}_{\textsc{bs}}} + \frac{1}{T_{\textsc{lt}}}\frac{\mathcal{C}_{\rm as}}{\mathcal{L}_{\textsc{bs}}},
\end{equation}
in which $\tau$ is the length of the uplink pilot signals, $\mathcal{S}$ is the coherence block size, $\mathcal{C}_{\rm ts}$ is the computational complexity {for} evaluating eq. \eqref{eq:transmit_sig}. Besides, $\mathcal{L}_{\textsc{bs}}$ is the computational efficiency of the BS (in W$/flop$), $\mathcal{C}_{\rm prec}$ is the complexity of obtaining the precoding vectors for all users, $T_{\textsc{lt}}$ is the long-term fading coherence time, and $\mathcal{C}_{\rm as}$ is the complexity of obtaining the antenna selection set. {The obtained {AS} set remains valid for a long-term coherence interval, since our analysis is based only in long-term fading parameters. One can see from \eqref{eq:pproc} that this approach results in a lower influence of the AS set computation in $P_{\textsc{pr}}$, since it is multiplied by the factor $1/T_{\textsc{lt}}$, which is much lower than $B/\mathcal{S}$ and $B \left( 1 - \frac{\tau}{\mathcal{S}} \right)$.}

Following the analysis in \cite{Marinello19}, \cite{Debbah15}, we consider 1 $flop$ as an arithmethic operation between two complex numbers. Thus, the multiplication between a matrix ${\bf A} \in \mathbb{C}^{m\times n}$ and a matrix ${\bf B} \in \mathbb{C}^{n\times p}$ spends $2 mnp$ flops. Therefore, we have $\mathcal{C}_{\rm ts} = 2 |\mathcal{A}| K \, flops$ from \cite{Debbah15}. Besides, if using the CB precoder, $\mathcal{C}_{\rm prec} = \mathcal{C}_{\rm CB} = 3 |\mathcal{A}| K \, flops$ from \cite{Debbah15}, against $\mathcal{C}_{\rm prec} = \mathcal{C}_{\rm ZF} = K^3/3 + 3 |\mathcal{A}| K^2 + |\mathcal{A}| K \, flops$ if adopting ZF. The complexity $\mathcal{C}_{\rm as}$ is discussed in the next Section. Besides, the terms in \eqref{eq:cp_model} not discussed in this Section can be computed in the same way as in \cite{Marinello19}.

{Finally, we can rewrite \eqref{eq:cp_model} as}
\begin{equation}\label{eq:cp_model2}
{\mathcal{P} = \mathcal{P}^\dagger +  P_{\textsc{ce}} + P_{\textsc{c/d}} + P_{\textsc{bh}} + P_{\textsc{pr}} + |\mathcal{A}| P_{\textsc{bs}},}
\end{equation}
{in which we have gathered the power components that do not depend of $\mathcal{A}$ in} {the term:}
\begin{equation}\label{eq:cp_model3}
{\mathcal{P}^\dagger = P_{\textsc{tx}}^{\textsc{dl}} + P_{\textsc{tx}}^{{\text{tr}}} + P_{\textsc{syn}} + K P_{\textsc{mt}} + P_{\textsc{fix}}.}
\end{equation}

The dependence of the terms in \eqref{eq:cp_model2} with $\mathcal{A}$ can be justified as follows: $P_{\textsc{ce}}$ depends on $\mathcal{A}$ since the short-term channel estimates are obtained only for the active antennas, $P_{\textsc{c/d}}$ and $P_{\textsc{bh}}$ because they depend on the system sum-rate, which depends on $\mathcal{A}$, and $P_{\textsc{pr}}$ because the processing complexity is dependent on the number of active antennas.

\section{Antenna Selection Schemes}\label{sec:contrib}
In this section we propose different AS schemes {for XL-MIMO aiming {to obtain}} a suitable subset of antennas $\mathcal{A}$ selected to transmit the DL signal to the {mobile users subject to channel non-stationarities}. First we propose a simple, deterministic, greedy scheme based on the {\it highest received normalized power} (HRNP) criterion. Then, three heuristic schemes are proposed using the HRNP active antennas set as initial solution: {\it local search} (LS), {\it genetic algorithm} (GA), and {\it particle swarm optimization} (PSO).

\subsection{HRNP criterion}
A first and greedy approach is to select just the $M_s$ antennas responsible for the major part of the power received by the users. However, since closer users receive more power, this should be performed in a normalized fashion in order to achieve a fair result for all users. In this case, we first compute the metric:
\begin{equation}\label{eq:hrnp}
\varphi_m = \sum_{k=1}^{K} \frac{\beta_{m,k}}{\sum_{j=1}^{M} \beta_{j,k}}, \quad m=1,2,\ldots,M.
\end{equation}
Then, the selected subset of antennas {$\mathcal{A}^{\textsc{hrnp}}$} will be composed by the $M_s$ antennas with the highest values of $\varphi_m$. {A pseudo-code for the HRNP-AS procedure is presented in Algorithm \ref{alg:HRNP}, in which $\bm{\varphi} = [\varphi_1, \varphi_2, \ldots \varphi_M]$.} 

The complexity\footnote{{We evaluate the computational complexities of the investigated schemes in terms of \emph{floating point operations (flops)}, defined as an addition, subtraction, multiplication or division between two floating point numbers \cite{Golub}.}} of the HRNP {AS} scheme {is described by}
\begin{equation}\label{eq:comphrnp}
\mathcal{C}^{\textsc{hrnp}}_{\rm as} = 3 M K + M \log(M) \,\,\,\, [flops],
\end{equation} 
corresponding to the computation of \eqref{eq:hrnp} for all antennas, and a sorting algorithm to select the $M_s$ antennas with highest $\varphi_m$. It is noteworthy, however, that the HRNP EE performance is highly dependent on the $M_s$ choice, since the system would provide low sum-rates with few active antennas, or it would consume a high power with many active antennas. Thus, we propose in Section \ref{sec:msopt} an approximated closed-form analytical expression for the EE of the XL-MIMO system employing ZF and HRNP-AS as a function of $M_s$. Then, we propose an iterative method for obtaining the $M_s$ value which maximizes this expression. We do not consider the complexity of this method in eq. \eqref{eq:comphrnp} since it is not dependent on the channel parameters, but only controlled by the system parameters, such as the number of users, transmit power, dimensions of XL-MIMO array and coverage area. Therefore, its computation can be performed over larger time periods. {We discuss in Section \ref{sec:compAS} the complexity of the proposed method for obtaining the optimal $M_s$ value.}

{\begin{algorithm}
\caption{\small{Proposed HRNP AS Scheme}}
\begin{flushleft}
Input: {$M_s$, $\beta_{m,k}$,  $\forall m, k$.}
\end{flushleft}
\label{alg:HRNP}
\begin{algorithmic}[1]
\STATE{Initialize $\mathcal{A}$ as an empty set;}
\FOR{$m = 1, 2, \ldots, M$}
    \STATE{Evaluate $\varphi_m$ as in \eqref{eq:hrnp};}
\ENDFOR	
\FOR{$n = 1, 2, \ldots, M_s$}
	\STATE{Evaluate $a = \arg \max_{m} \varphi_m$;}
	\STATE{Update $\mathcal{A}$ as $\mathcal{A} = \mathcal{A} \cup a$;} 
	\STATE{Remove $\varphi_a$ from $\bm{\varphi}$;}
\ENDFOR
\end{algorithmic}
Output: $\mathcal{A}^{\textsc{hrnp}}$.
\end{algorithm}}
\normalsize

\subsection{LS-based Antenna Selection}
A simple strategy for seeking a better active antennas set is to perform a local search (LS) in the neighborhood of the HRNP solution. For this purpose, {we first represent} the set $\mathcal{A}$ as a binary vector ${\bf{a}}$ of length $M$, {in which if $m \in \mathcal{A}$, $a_m = 1$; otherwise $a_m = 0$. Then}, we compute the total energy efficiency \eqref{eq:eneff} of every candidate within a certain Hamming distance {$d_{\rm{Ham}}$} from it. If a better candidate is found, the solution is updated, and the procedure is repeated on its neighborhood. This iterative procedure is repeated for a predefined number of iterations or until the convergence. {A pseudo-code representation of the LS-based AS scheme is provided in Algorithm \ref{alg:LS}, {in which $N_{el}$ as defined in step 3 is the number of elements within the Hamming distance {$d_{\rm{Ham}}$} from the current solution}.} For simplicity, we have limited our search with a unitary Hamming distance. 

The complexity of the LS-AS scheme is 
\begin{equation}\label{eq:compls}
\mathcal{C}^{\textsc{ls}}_{\rm as} = \mathcal{C}^{\textsc{hrnp}}_{\rm as} + \overline{N}_{it} M \mathcal{C}_{\textsc{ee}} \,\,\,\,\, [flops],
\end{equation}
in which $\overline{N}_{it}$ is the average number of iterations until convergence, and $\mathcal{C}_{\textsc{ee}} = 2 M K^2 \, [flops]$ is the complexity of computing the total energy efficiency cost function. An interesting point to observe in the LS algorithm is that if a new solution is not found into an iteration, the search can be interrupted, since the algorithm has converged. This contributes to decrease the complexity of the algorithm, and, therefore, improve EE.

{\begin{algorithm}
\caption{\small{Proposed LS-based AS Scheme}}
\begin{flushleft}
Input: {$d_{\rm{Ham}}$, $N_{it}^{\max}$, $\mathcal{A}^{\textsc{hrnp}}$, $\beta_{m,k}$,  $\forall m, k$.}
\end{flushleft}
\label{alg:LS}
\begin{algorithmic}[1]
\STATE{Initialize ${\bf a}^0$ as the binary vector representation of $\mathcal{A}^{\textsc{hrnp}}$;}
\STATE{Initialize $\eta_e^{\rm best}$ as the total energy efficiency of ${\bf a}^0$;}
\STATE{Evaluate $N_{el} = {{M}\choose{d_{\rm{Ham}}}}$;}
\FOR{$n = 1, 2, \ldots, N^{{\rm max}}_{it}$}
	\STATE{Generate the search space matrix $\bf S$ of size $M \times N_{el}$ with all vectors within the distance $d_{\rm{Ham}}$ from ${\bf a}^{n-1}$;}
	\FOR{$\ell = 1, 2, \ldots, N_{el}$}
	    \STATE{Evaluate $\eta_e$ as the total energy efficiency of $\bf{s}_{\ell}$;}
	    \IF{$\eta_e>\eta_e^{\rm best}$}
	        \STATE{Update $\eta_e^{\rm best} = \eta_e$, and ${\bf a}^{n}=\bf{s}_{\ell}$;}
	    \ELSE
	       \STATE{\textbf{Break;}}
	    \ENDIF
	\ENDFOR
\ENDFOR
\end{algorithmic}
Output: $\mathcal{A}^{\textsc{ls}}$ as the set representation of ${\bf a}^{n}$.
\end{algorithm}}
\normalsize

\subsection{GA-based Antenna Selection}\label{sec:GA}
The genetic algorithm is a widely-known bio-inspired heuristic optimization algorithm, which has been used to solve optimization problems in different areas. In the context of massive MIMO antenna selection, GA has been employed in the conventional stationary case in \cite{Alouini17}. Herein, we employ a similar algorithm {from \cite{Alouini17}, but adjusted to} the non-stationary XL-MIMO {configurations}. The GA-AS uses the HRNP output as initial solution, {also,} other random candidates forming an initial population of size $p_{\textsc{ga}}$, which is evaluated in terms of the cost function in \eqref{eq:eneff}. A given {number $\phi$} of the best candidates in this population is selected as parents, which will generate descendants in a new population. For this purpose, two parents are selected at random for each descendant, and the crossover operator is applied with a random crossover point. Then, the mutation operator is also applied, which inverts the entries of each candidate with certain probability {$p_{\rm mut}$}. After a predefined number of iterations or until the convergence of the algorithm, it returns the best solution found so far. {A pseudo-code representation for the GA-based AS scheme is provided in Algorithm \ref{alg:GA}.}

The complexity of our proposed GA-AS procedure is 
\begin{equation}\label{eq:compga}
\mathcal{C}^{\textsc{ga}}_{\rm as} = \mathcal{C}^{\textsc{hrnp}}_{\rm as} + \overline{N}_{it}[p_{\textsc{ga}} \mathcal{C}_{\textsc{ee}} + p_{\textsc{ga}} \log(p_{\textsc{ga}})] \,\,\,\,\, [flops],
\end{equation}
due to the cost function evaluation of each candidate in the population, and a sorting algorithm for selecting the best candidates.

{\begin{algorithm}
\caption{\small{Proposed GA-based AS Scheme}}
\begin{flushleft}
Input: {$p_{\textsc{ga}}$, $\phi$, $p_{\rm mut}$, $\mathcal{A}^{\textsc{hrnp}}$, $N_{it}^{\max}$,  $\beta_{m,k}$,  $\forall m, k$.}
\end{flushleft}
\label{alg:GA}
\begin{algorithmic}[1]
\STATE{Initialize the population $\bm{\Theta}^{\textsc{ga}}$ with the binary vector representation of $\mathcal{A}^{\textsc{hrnp}}$ and other $p_{\textsc{ga}}$-1 random binary vectors;}
\STATE{Evaluate the total energy efficiency of each candidate in $\bm{\Theta}^{\textsc{ga}}$, forming the vector $\bm{\eta}_e^{\textsc{ga}}$;}
\STATE{Sort $\bm{\eta}_e^{\textsc{ga}}$ in descending order, reorganizing the columns of $\bm{\Theta}^{\textsc{ga}}$ accordingly;}
\STATE{Initialize $\eta_e^{\rm best}={\eta}_{e,1}^{\textsc{ga}}$, and ${\bf a}^{\textsc{ga}}=\bm{\theta}^{\textsc{ga}}_1$;}
\FOR{$n = 2, 3, \ldots, N^{{\rm max}}_{it}$}
	\FOR{$\ell = 1, 2, \ldots, p_{\textsc{ga}}$}
	    \STATE{Generate two different random integers $\in [1, \phi]$ to be the parents of $\bm{\theta}^{\textsc{ga}}_{\ell}$, applying the crossover operator in a random crossover point $\in [2, M]$;}
	    \STATE{Apply the mutation operator in $\bm{\theta}^{\textsc{ga}}_{\ell}$ with probability $p_{\rm mut}$;}
	    \STATE{Evaluate the total energy efficiency of $\bm{\theta}^{\textsc{ga}}_{\ell}$, and assign it to ${\eta}_{e,\ell}^{\textsc{ga}}$;}
	\ENDFOR
	\STATE{Sort $\bm{\eta}_e^{\textsc{ga}}$ in descending order, reorganizing the columns of $\bm{\Theta}^{\textsc{ga}}$ accordingly;}
    \IF{${\eta}_{e,1}^{\textsc{ga}}>\eta_e^{\rm best}$}
	    \STATE{Update $\eta_e^{\rm best} = {\eta}_{e,1}^{\textsc{ga}}$, and ${\bf a}^{\textsc{ga}}=\bm{\theta}^{\textsc{ga}}_1$;}
	\ENDIF
\ENDFOR
\end{algorithmic}
Output: $\mathcal{A}^{\textsc{ga}}$ as the set representation of ${\bf a}^{\textsc{ga}}$.
\end{algorithm}}
\normalsize

\subsection{PSO-based Antenna Selection}
The particle swarm optimization algorithm is another bio-inspired optimization algorithm, similarly as GA. However, it is commonly recognized as a simpler algorithm, in terms of fewer mechanisms to escape from local maxima, and reduced computational complexity per iteration. Therefore, we also {suggest the use of} a PSO-based {AS} scheme for the non-stationary XL-MIMO case, similarly as proposed in \cite{Sun17} {for} conventional stationary massive MIMO scenario.

The PSO-AS algorithm uses the HRNP output as initial solution, as well as other random candidates to form an initial swarm of $p_{\textsc{pso}}$ particles. At each iteration, each particle updates its position in terms of its previous velocity (inertial effect, {with inertia weight $\nu$}), its individual best solution found (cognitive information, {with cognitive factor $\mu_c$}), and the best solution found by all particles (social information, {with social factor $\mu_s$}). After a predefined number of iterations or the convergence of the algorithm, it returns the best solution found. {A pseudo-code representation for the PSO-based AS scheme is provided in Algorithm \ref{alg:PSO}, in which $\bm{\Gamma} \in \mathbb{R}^{M\times p_{\textsc{pso}}}$ is a random matrix generated each time it is called with each element uniformly distributed in $[0,1]$ interval, and ${\rm \texttt{binround}}(x)$ is the binary round operator, which returns 1 if $x>0.5$, and 0 otherwise.}

The complexity of the proposed PSO-AS algorithm is 
\begin{equation}\label{eq:comppso}
\mathcal{C}^{\textsc{pso}}_{\rm as} = \mathcal{C}^{\textsc{hrnp}}_{\rm as} + \overline{N}_{it}(p_{\textsc{pso}} \mathcal{C}_{\textsc{ee}} + p_{\textsc{pso}}) \,\,\,\, [flops],
\end{equation}
due to the cost function evaluation \eqref{eq:eneff} for all particles and finding the maximum EE particle, at each iteration.

{\begin{algorithm}
\caption{\small{Proposed PSO-based AS Scheme}}
\begin{flushleft}
Input: {$p_{\textsc{pso}}$, $\nu$, $\mu_c$, $\mu_s$, $\mathcal{A}^{\textsc{hrnp}}$, $N_{it}^{\max}$,  $\beta_{m,k}$,  $\forall m, k$.}
\end{flushleft}
\label{alg:PSO}
\begin{algorithmic}[1]
\STATE{Initialize the positions $\bm{\Theta}^{\textsc{pso}}$ with the binary vector representation of $\mathcal{A}^{\textsc{hrnp}}$ and other $p_{\textsc{pso}}$-1 random binary vectors;}
\STATE{Evaluate the total energy efficiency of each candidate in $\bm{\Theta}^{\textsc{pso}}$, forming the vector $\bm{\eta}_e^{\textsc{pso}}$;}
\STATE{Initialize the social information $\eta_e^{\rm best}={\eta}_{e,\phi}^{\textsc{pso}}$, and ${\bf a}^{\textsc{pso}}=\bm{\theta}^{\textsc{pso}}_{\phi}$, in which $\phi=\arg \max_n {\eta}_{e,n}^{\textsc{pso}}$;}
\STATE{Initialize the cognitive information $\bm{\eta}_e^c=\bm{\eta}_e^{\textsc{pso}}$, and $\bm{\Theta}_c^{\textsc{pso}}=\bm{\Theta}^{\textsc{pso}}$;}
\STATE{Initialize the velocity matrix ${\bf V} \in \mathbb{R}^{M\times p_{\textsc{pso}}}$ with random elements uniformly distributed in $[-1,1]$;}
\FOR{$n = 2, 3, \ldots, N^{\max}_{it}$}
    \STATE{Update the velocity matrix \break {${\bf V} = \nu {\bf V} + \mu_c \bm{\Gamma} \left[\bm{\Theta}_c^{\textsc{pso}} - \bm{\Theta}^{\textsc{pso}} \right] + \mu_s \bm{\Gamma} \left[{\bf a}^{\textsc{pso}} - \bm{\Theta}^{\textsc{pso}} \right]$};}
	\STATE{Update the positions $\bm{\Theta}^{\textsc{pso}} = {\rm \texttt{binround}}\left(\bm{\Theta}^{\textsc{pso}} + {\bf V} \right)$;}
	\FOR{$\ell = 1,2,\ldots p_{\textsc{pso}}$}
\STATE{Evaluate the total energy efficiency of $\bm{\theta}_{\ell}^{\textsc{pso}}$, and assign it to ${\eta}_{e,\ell}^{\textsc{pso}}$;}
\IF{${\eta}_{e,\ell}^{\textsc{pso}}>{\eta}_{e,\ell}^c$}
\STATE{Update ${\eta}_{e,\ell}^c = {\eta}_{e,\ell}^{\textsc{pso}}$, and $\bm{\theta}_{c,\ell}^{\textsc{pso}}=\bm{\theta}^{\textsc{pso}}_{\ell}$;}
	        \IF{${\eta}_{e,\ell}^{\textsc{pso}}>\eta_e^{\rm best}$}
	            \STATE{Update $\eta_e^{\rm best} = {\eta}_{e,\ell}^{\textsc{pso}}$, and ${\bf a}^{\textsc{pso}}=\bm{\theta}^{\textsc{pso}}_{\ell}$;}
	        \ENDIF
	    \ENDIF
	\ENDFOR
\ENDFOR
\end{algorithmic}
Output: $\mathcal{A}^{\textsc{pso}}$ as the set representation of ${\bf a}^{\textsc{pso}}$.
\end{algorithm}}
\normalsize

\section{Optimal number of selected antennas: an iterative-analytical method }\label{sec:msopt}

In this Section we derive approximated performance analytical expressions for XL-MIMO systems employing the ZF precoder and the HRNP-based AS method. Such expressions are compared with numerical results obtained via Monte-Carlo simulation method in Section \ref{sec:results}, confirming the tightness of the derivations proposed herein. Then, based on these analytical expressions, we devise an analytical iterative algorithm based on Newton-Raphson (NR) method to determine the optimal number of activated antennas for XL-MIMO systems, which maximizes the approximated EE expression.

In order to compute the average ZF SINR expression, one can directly evaluate from eq. \eqref{eq:SINR_ZF_as}:
{\small
\begin{eqnarray}\label{eq:SINR_ZF_as2}
\mathbb{E}\left[\overline{\gamma}_k^{(\textsc{ZF})}\right] = \mathbb{E}\left[\frac{P_{\rm max}}{K \sigma^2} \left(\sum_{m \in \mathcal{A}} \beta_{m,k} - \sum_{j=1, j\neq k}^{K} \frac{\sum_{m \in \mathcal{A}} \beta_{m,k} \beta_{m,j}}{\sum_{m \in \mathcal{A}} \beta_{m,j}} \right)\right]\nonumber\\
= \mathbb{E}\left[\frac{P_{\rm max}}{K \sigma^2} \left(\sum_{m \in \mathcal{A}} \beta_{m,k} - \sum_{m \in \mathcal{A}} \beta_{m,k} \sum_{j=1, j\neq k}^{K} \frac{\beta_{m,j}}{\sum_{n \in \mathcal{A}} \beta_{n,j}} \right)\right]\nonumber\\
= \mathbb{E}\left[\frac{P_{\rm max}}{K \sigma^2} \sum_{m \in \mathcal{A}} \beta_{m,k} \left(1 - \sum_{j=1, j\neq k}^{K} \frac{\beta_{m,j}}{\sum_{n \in \mathcal{A}} \beta_{n,j}} \right)\right]\nonumber\\
= \frac{P_{\rm max}}{K \sigma^2} \sum_{m \in \mathcal{A}} \mathbb{E}\left[\beta_{m,k}\right] \left(1 - \sum_{j=1, j\neq k}^{K} \mathbb{E}\left[\frac{\beta_{m,j}}{\sum_{n \in \mathcal{A}} \beta_{n,j}}\right] \right)
\end{eqnarray}
}
\normalsize
in which the expectation is taken with respect to the random users' positions. 

Instead of advancing with \eqref{eq:SINR_ZF_as2} seeking an exact solution, we approximate the average SINR by the SINR of a {{\it user in the most expected position} (\textsc{umep})}. Given the uniform distribution of the users {as illustrated in Fig.} \ref{fig:SystModel}, this most expected position would be as depicted in Fig. \ref{fig:SystModel_med}.

\begin{figure}[htbp]
\centering
\includegraphics[width=0.464\textwidth]{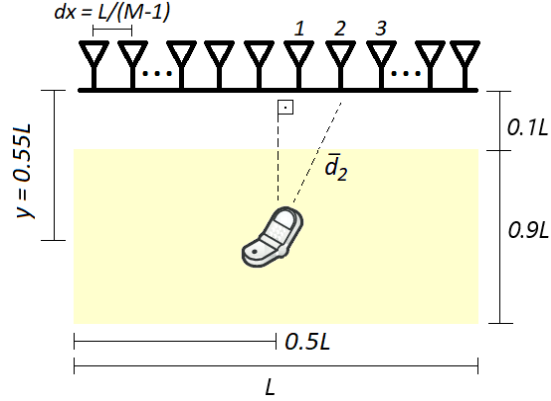} 
\vspace{-2mm}
\caption{Illustration of the most expected user's position {(\textsc{umep})}.}
\label{fig:SystModel_med}
\end{figure} 
\vspace{-2mm}

Then, considering this position for the users, and noting that the HRNP AS activate in this case the $M_s$ closest antennas, the ZF SINR expression becomes
\begin{eqnarray}\label{eq:SINR_ZF_as3}
\mathbb{E}\left[\overline{\gamma}_k^{(\textsc{ZF})}\right] \approx \frac{P_{\rm max}}{K \sigma^2} \left(\sum_{m \in \mathcal{A}} \overline{\beta}_{m} - (K-1) \frac{\sum_{m \in \mathcal{A}} \overline{\beta}_{m}^2}{\sum_{m \in \mathcal{A}} \overline{\beta}_{m}} \right),\nonumber\\
\approx \frac{P_{\rm max}}{K \sigma^2} \left(2\sum_{m =1}^{M_s/2} \overline{\beta}_{m} - (K-1) \frac{2\sum_{m =1}^{M_s/2} \overline{\beta}_{m}^2}{2\sum_{m =1}^{M_s/2} \overline{\beta}_{m}} \right),
\end{eqnarray}
with $\overline{\beta}_{m} = q \cdot (\overline{d}_{m})^{-\kappa}$, and $\overline{d}_{m} = \sqrt{y^2 + [(m-\frac{1}{2})dx]^2} = y\sqrt{1 + [(m-\frac{1}{2})\frac{dx}{y}]^2} \approx y\sqrt{1 + (m\frac{dx}{y})^2}$ is represented in {Fig.} \ref{fig:SystModel_med} for $m=2$. Eq. \eqref{eq:SINR_ZF_as3} can thus be simplified as in the next page, in which from \eqref{eq:SINR_ZF_as4} to \eqref{eq:SINR_ZF_as5} we have used the binomial approximation $(1+x)^{\alpha}\approx 1+\alpha x$ for $|\alpha x|\ll 1$. In our scenario, this condition becomes 
\begin{equation}\label{eq:cond_approx}
\frac{\kappa \, M_s^2 \, dx^2}{8 \, y^2} \ll 1,    
\end{equation}
which usually holds for typical XL-MIMO systems. For example, the binomial approximation results in relative errors lower than $5\%$ for $|\alpha x|<0.25$, which in our XL-MIMO scenario corresponds to $M_s<225$. Besides, with this approximated ZF HRNP-AS SINR expression, we can also approximate the EE {expression} as in eq. \eqref{eq:eneff2}. {Moreover, by} expanding all the power terms in {the denominator of} \eqref{eq:eneff2}, as discussed in Section \ref{sec:power}, and grouping them according to their dependence with $M_s$, we arrive at eq. \eqref{eq:eneff3}, in which $\mathcal{P}_{\textsc{bc}} = \mathcal{P}_{\textsc{cod}}+\mathcal{P}_{\textsc{dec}}+\mathcal{P}_{\textsc{bt}}$, and $\mathcal{T}_{0}$ $,\mathcal{T}_{1}$ {are} defined in eq. \eqref{eq:T0} and \eqref{eq:T1}, respectively.
\small
\begin{align}
\eta_e \approx \frac{B K \, \log_2\left(1+\mathbb{E}\left[\overline{\gamma}_k^{(\textsc{ZF})}\right]\right)}{P_{\textsc{tx}}^{\textsc{dl}} + P_{\textsc{tx}}^{{\text{tr}}} + P_{\textsc{ce}} + P_{\textsc{c/d}} + P_{\textsc{bh}} + P_{\textsc{pr}} + P_{\textsc{tc}} + P_{\textsc{fix}}},\label{eq:eneff2}\\
\approx \frac{B K \, \log_2\left(1+\mathbb{E}\left[\overline{\gamma}_k^{(\textsc{ZF})}\right]\right)}{\mathcal{P}_{\textsc{bc}} B K \, \log_2\left(1+\mathbb{E}\left[\overline{\gamma}_k^{(\textsc{ZF})}\right]\right) + \mathcal{T}_0 + \mathcal{T}_1 M_s}.\label{eq:eneff3}
\end{align} 
\begin{equation}\label{eq:T0}
{\mathcal{T}_0 = \mathcal{P}^\dagger + \frac{3 M K + M \log(M)}{T_{\textsc{lt}} \, \mathcal{L}_{\textsc{bs}}} + \frac{B \, K^3}{3 \, \mathcal{S} \, \mathcal{L}_{\textsc{bs}}}.}
\end{equation}
\begin{equation}\label{eq:T1}
{\mathcal{T}_1 = \mathcal{P}_{\textsc{bs}} + \frac{5 \, B \, K^2}{\mathcal{S} \, \mathcal{L}_{\textsc{bs}}} + \left( 1 - \frac{\tau}{\mathcal{S}} \right) \frac{2 \, B \, K}{\mathcal{L}_{\textsc{bs}}} + \frac{B \, K}{\mathcal{S} \, \mathcal{L}_{\textsc{bs}}}.}
\end{equation}
\normalsize

\begin{figure*}[!ht]
\footnotesize
\begin{flalign}
\mathbb{E}\left[\overline{\gamma}_k^{(\textsc{ZF})}\right] &\approx \frac{P_{\rm max} q}{K \sigma^2} \left(2 \sum_{m=1}^{M_s/2} (\overline{d}_{m})^{-\kappa} - (K-1) \frac{2 \sum_{m=1}^{M_s/2} (\overline{d}_{m})^{-2\kappa}}{2 \sum_{m=1}^{M_s/2} (\overline{d}_{m})^{-\kappa}} \right),\nonumber\\
&\approx \frac{P_{\rm max} q}{K \sigma^2} \left\{2 \sum_{m=1}^{M_s/2} y^{-\kappa}\left[1 + \left(\frac{m \, dx}{y}\right)^2\right]^{-\frac{\kappa}{2}} - (K-1) \frac{2\sum_{m=1}^{M_s/2} y^{-2\kappa}\left[1 + \left(\frac{m \, dx}{y}\right)^2\right]^{-\kappa}}{2\sum_{m=1}^{M_s/2} y^{-\kappa}\left[1 + \left(\frac{m \, dx}{y}\right)^2\right]^{-\frac{\kappa}{2}}} \right\},\nonumber\\
&\approx \frac{P_{\rm max} q \, y^{-\kappa}}{K \sigma^2} \left\{2 \sum_{m=1}^{M_s/2} \left[1 + \left(\frac{m \, dx}{y}\right)^2\right]^{-\frac{\kappa}{2}} - (K-1) \frac{2\sum_{m=1}^{M_s/2} \left[1 + \left(\frac{m \, dx}{y}\right)^2\right]^{-\kappa}}{2\sum_{m=1}^{M_s/2} \left[1 + \left(\frac{m \, dx}{y}\right)^2\right]^{-\frac{\kappa}{2}}} \right\} \qquad {(ZF_{ME})},\label{eq:SINR_ZF_as4}\\
&\approx \frac{P_{\rm max} q \, y^{-\kappa}}{K \sigma^2} \left\{2 \sum_{m=1}^{M_s/2} \left[1 - \frac{\kappa}{2} \left(\frac{m \, dx}{y}\right)^2\right] - (K-1) \frac{2\sum_{m=1}^{M_s/2} \left[1 - \kappa\left(\frac{m \, dx}{y}\right)^2\right]}{2\sum_{m=1}^{M_s/2} \left[1 -\frac{\kappa}{2} \left(\frac{m \, dx}{y}\right)^2\right]} \right\}, \label{eq:SINR_ZF_as5}\\
&\approx \frac{P_{\rm max} q \, y^{-\kappa}}{K \sigma^2} \left[\left(\mathcal{T}_{1,1} M_s - \mathcal{T}_{1,2} M_s^2 - \mathcal{T}_{1,3} M_s^3\right) - (K-1) \frac{\left(\mathcal{T}_{2,1} M_s - \mathcal{T}_{2,2} M_s^2 - \mathcal{T}_{2,3} M_s^3\right)}{\left(\mathcal{T}_{1,1} M_s - \mathcal{T}_{1,2} M_s^2 - \mathcal{T}_{1,3} M_s^3\right)} \right] \qquad {(ZF_{BA})},\label{eq:SINR_ZF_as6}\\
{\rm with} \quad \mathcal{T}_{1,1}&= 1-\frac{K \, dx^2}{12 \, y^2}, \quad \mathcal{T}_{1,2}=\frac{K \, dx^2}{8 \, y^2}, \quad
\mathcal{T}_{1,3}=\frac{K \, dx^2}{24 \, y^2}, \quad
\mathcal{T}_{2,1}=1-\frac{K \, dx^2}{6 \, y^2}, \quad \mathcal{T}_{2,2}=\frac{K \, dx^2}{4 \, y^2}, \quad
\mathcal{T}_{2,3}=\frac{K \, dx^2}{12 \, y^2}.\nonumber
\end{flalign}
\end{figure*}
\normalsize

\begin{figure*}
\scriptsize
\begin{flalign}
\frac{\partial f(M_s)}{\partial M_s} &= \frac{\partial^2 \mathbb{E}\left[\overline{\gamma}_k^{(\textsc{ZF})}\right]}{\partial M_s^2} - \frac{\mathcal{T}_1^2 \ln(2) \left(1+\mathbb{E}\left[\overline{\gamma}_k^{(\textsc{ZF})}\right]\right) \log_2\left(1+\mathbb{E}\left[\overline{\gamma}_k^{(\textsc{ZF})}\right]\right) - \mathcal{T}_1\left(\mathcal{T}_0 + \mathcal{T}_1 M_s\right) \frac{\partial \mathbb{E}\left[\overline{\gamma}_k^{(\textsc{ZF})}\right]}{\partial M_s} \left( 1 + \ln(2) \log_2\left(1+\mathbb{E}\left[\overline{\gamma}_k^{(\textsc{ZF})}\right]\right) \right)}{\left(\mathcal{T}_0 + \mathcal{T}_1 M_s\right)^2},\label{eq:derfMs}\\
{\rm with} &\quad \frac{\partial \mathbb{E}\left[\overline{\gamma}_k^{(\textsc{ZF})}\right]}{\partial M_s} = \frac{P_{\rm max} q \, y^{-\kappa}}{K \sigma^2} \left[\mathcal{F}_1' - 
(K-1)\frac{\mathcal{F}_1 \mathcal{F}_2' -\mathcal{F}_2 \mathcal{F}_1'}{\mathcal{F}_1^2} \right],\label{eq:dergzf}\\
{\rm and} &\quad \frac{\partial^2 \mathbb{E}\left[\overline{\gamma}_k^{(\textsc{ZF})}\right]}{\partial M_s^2} = \frac{P_{\rm max} q \, y^{-\kappa}}{K \sigma^2} \left[\mathcal{F}_1'' - 
(K-1) \frac{\mathcal{F}_1^2\left(\mathcal{F}_1 \mathcal{F}_2'' -\mathcal{F}_2 \mathcal{F}_1'' \right)-2\left(\mathcal{F}_1 \mathcal{F}_2' - \mathcal{F}_2 \mathcal{F}_1' \right) \mathcal{F}_1 \mathcal{F}_1'}{\mathcal{F}_1^4}\right],\label{eq:der2gzf}\\
{\rm in}\,\,{\rm  which}\,\, & \mathcal{F}_1 = \mathcal{T}_{1,1} M_s - \mathcal{T}_{1,2} M_s^2 - \mathcal{T}_{1,3} M_s^3, \quad \mathcal{F}_2 = \mathcal{T}_{2,1} M_s - \mathcal{T}_{2,2} M_s^2 - \mathcal{T}_{2,3} M_s^3, \nonumber\\
&\mathcal{F}_1' = \mathcal{T}_{1,1} - 2 \mathcal{T}_{1,2} M_s - 3 \mathcal{T}_{1,3} M_s^2, \quad \mathcal{F}_2' = \mathcal{T}_{2,1} - 2 \mathcal{T}_{2,2} M_s - 3 \mathcal{T}_{2,3} M_s^2, \quad \mathcal{F}_1'' = - 2 \mathcal{T}_{1,2} - 6 \mathcal{T}_{1,3} M_s, \quad \mathcal{F}_2'' = - 2 \mathcal{T}_{2,2} - 6 \mathcal{T}_{2,3} M_s.\nonumber
\end{flalign}
\end{figure*}
\normalsize

\subsection{Optimal Number of Activated Antennas}\label{sec:Msopt}
Considering our previous analytical results, we propose in this Section a method for obtaining the optimal $M_s$ value when employing ZF with HRNP AS, by taking the derivative of eq. \eqref{eq:eneff3}, with the SINR given in eq. \eqref{eq:SINR_ZF_as6}, with respect to $M_s$, and equaling it to 0 when $M_s = M_s^*$. Following this procedure, {and} after some simplifications, we arrive at $f(M_s^*) = 0$, with $f(M_s)$ defined as
\small
\begin{equation}\label{eq:Msf}
     f(M_s) = \frac{\partial \mathbb{E}\left[\overline{\gamma}_k^{(\textsc{ZF})}\right]}{\partial M_s} - \frac{\mathcal{T}_1 \ln(2) \left(1+\mathbb{E}\left[\overline{\gamma}_k^{(\textsc{ZF})}\right]\right) \log_2\left(1+\mathbb{E}\left[\overline{\gamma}_k^{(\textsc{ZF})}\right]\right)}{\mathcal{T}_0 + \mathcal{T}_1 M_s},
\end{equation}
\normalsize
where $\frac{\partial \mathbb{E}\left[\overline{\gamma}_k^{(\textsc{ZF})}\right]}{\partial M_s}$ is given in \eqref{eq:dergzf}.

Since $\mathbb{E}\left[\overline{\gamma}_k^{(\textsc{ZF})}\right]$ and its derivative are dependent of $M_s$, we cannot arrive at a closed-form expression for $M_s^*$. However, we can find the root of $f(M_s)$ by applying some iterative numerical method, like Newton-Raphson (NR) method, which obtains a sequence of $M_s$ values $M_{s,0}, M_{s,1}, M_{s,2}, \ldots M_{s,n}$ converging to $M_s^*$ if the starting point $M_{s,0}$ is not too far from it. The values in the sequence {obey}
\begin{equation}\label{eq:Msn}
    M_{s,n} = M_{s,n-1} - \frac{f(M_{s,n-1})}{\left. \frac{\partial f(M_s)}{\partial M_s}\right|_{M_{s,n-1}}},
\end{equation}
in which the derivative of $f(M_s)$ is given in \eqref{eq:derfMs}.

\section{Numerical Results and Discussion}\label{sec:results}
Our adopted simulation parameters are indicated in Table \ref{tab:parameters}. While we have chosen very similar power consumption parameters than that of \cite{Marinello19}, \cite{Debbah15}, the XL-MIMO system parameters are chosen similarly as \cite{amiri2020deep, yang2019uplink, Heath19}, as well as in accordance with common XL-MIMO scenario applications. Considering $M = 512$ antennas at the XL-MIMO BS, Fig. \ref{fig:SE_EE} {depicts} the SINR, sum {SE} and the energy efficiency as a function of {number of users} $K$ (from 1 to $M/2$), for both CB and ZF precoders. {The sum SE is presented in units of bits per channel use (bpcu).} One can note that ZF precoding always achieve a higher total energy efficiency than CB in the scenario investigated. The presented results were averaged among {1000} random realizations of the users' positions. It is also shown in the Figure the equivalence between the results of performance expressions from \cite{Heath19}, eq. \eqref{eq:SINR_CB_de} and \eqref{eq:SINR_ZF_de}, and the expressions with our proposed simplifications, eq. \eqref{eq:SINR_CB_de2} and \eqref{eq:SINR_ZF_de2}.

\begin{table}[!htbp]
\caption{Simulation Parameters.}
\vspace{-2mm}
\centering
\small
{\renewcommand{\arraystretch}{1.1}%
\begin{tabular}{rl}
\hline
\bf Parameter & \bf Value\\
\hline
Carrier frequency: $f$ & 2.6 GHz\\
{Number of BS antennas $M$} & {$[500; \, 512]$}\\
XL-MIMO array length: $L$ & 30 m\\
Distance of users to BS: & $[0.1 \cdot L, L]$\\
Path loss decay exponent: $\kappa$ & 3\\
Path loss at the reference distance: $q$ & $10^{-3.53}$\\
Transmission bandwidth: $B$ & 20 MHz\\
Channel coherence bandwidth: $B_C$ & 100 kHz\\
Channel coherence time: $T_C$ & 2 ms\\
Long-term fading coherence time: $T_{\textsc{lt}}$ & 2 s\\
Total noise power: $\sigma^2$ & $-96$ dBm\\
UL pilot transmit power: $\rho_{\rm p}$ & 20 mW\\
DL radiated power: $P_{\rm max}=\frac{\rho \sigma^2}{q L^{-\kappa}}$ & 0.23 mW\\
Coherence block: $\mathcal{S}$ & 200 symbols\\
Length of the uplink pilot signals: $\tau$ & $K$\\
Computational efficiency at BSs: $\mathcal{L}_{\textsc{bs}}$ & $12.8 \, \left[\frac{\rm Gflops}{\rm W}\right]$\\
Fraction of DL transmission: $\xi^{\rm d}$ & 1\\
Fraction of UL transmission: $\xi^{\rm u}$ & 0\\
PA efficiency at the BS: $\eta^{\rm d}$ & 0.39\\
PA efficiency at the MTs: $\eta^{{\rm u} \textsc{t}}$ & 0.50\\
Fixed power consumption: $P_{\textsc{fix}}$ & 18 W\\
Power for local oscillators at BSs: $P_{\textsc{syn}}$ & 2 W\\
Power for circuit components BSs: $P_{\textsc{bs}}$ & 1 W\\
Power for circuit components MTs: $P_{\textsc{mt}}$ & 0.10 W\\
Power density for coding data: $\mathcal{P}_{\textsc{cod}}$ & $0.10 \, \left[\frac{\rm W}{\rm Gbit/s}\right]$\\
Power density for decoding data: $\mathcal{P}_{\textsc{dec}}$ & $0.80 \, \left[\frac{\rm W}{\rm Gbit/s}\right]$\\
Power density for backhaul traffic: $\mathcal{P}_{\textsc{bt}}$ & $0.25 \, \left[\frac{\rm W}{\rm Gbit/s}\right]$\\
 \hline
\end{tabular}}
\label{tab:parameters}
\end{table}

\begin{figure}[!htbp]
\centering
\includegraphics[width=.85\textwidth]{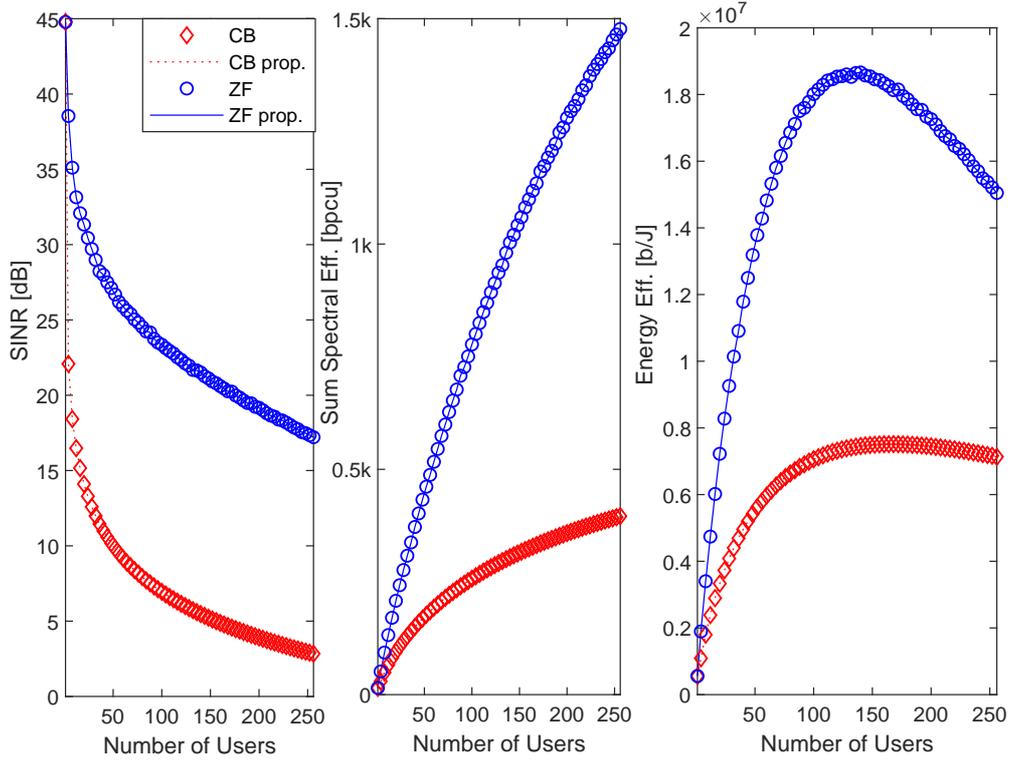} \\
\vspace{-4mm}
\caption{(a) SINR, (b) sum-SE, and (c) EE {\it vs.} $K$ for $M = 512$ antennas, selecting all available antennas. Proposed eq. \eqref{eq:SINR_CB_de2} and \eqref{eq:SINR_ZF_de2}, are represented by dotted and solid line curves, respectively, while the performances from \cite{Heath19}, eq. \eqref{eq:SINR_CB_de} and \eqref{eq:SINR_ZF_de}, are indicated by the curves with '$\Diamond$' and 'o' markers.} 
\label{fig:SE_EE}
\end{figure}

Now, considering $M = 500$ antennas at the XL-MIMO BS, and the same power consumption parameters, Fig. \ref{fig:SE_EE_selant} shows the SINR, sum SE and the EE as a function of $M_s\in \{100;\, M\}$, with $K=100$ users, for both CB and ZF precoders when employing the HRNP {AS} scheme. Notice that ZF precoding achieves a higher total energy efficiency than CB in the scenario investigated. Besides, {by} activating a number of $M_s = 146$ BS antennas, one can attain the maximum total energy efficiency for ZF precoder with $K=100$ users ("$M_s^*$ by NR" point in Fig. \ref{fig:SE_EE_selant}.c), as found by our proposed NR method of Section \ref{sec:Msopt}. Fig. \ref{fig:SE_EE_selant} also compares the performance obtained by averaging eq. \eqref{eq:SINR_ZF_as} with several random realizations for the users' positions (denoted as ZF), with the approximated deterministic result from eq. \eqref{eq:SINR_ZF_as4}, denoted as ZF$_{ME}$, and with the binomial approximation in eq. \eqref{eq:SINR_ZF_as6}, denoted as ZF$_{BA}$. It also shows the results in terms of sum SE and EE of the XL-MIMO system. One can conclude that both proposed approximations are tight, and that the $M_s$ values that maximize them are nearly the same.

\begin{figure}[!htbp]
\centering
\includegraphics[width=.85\textwidth]{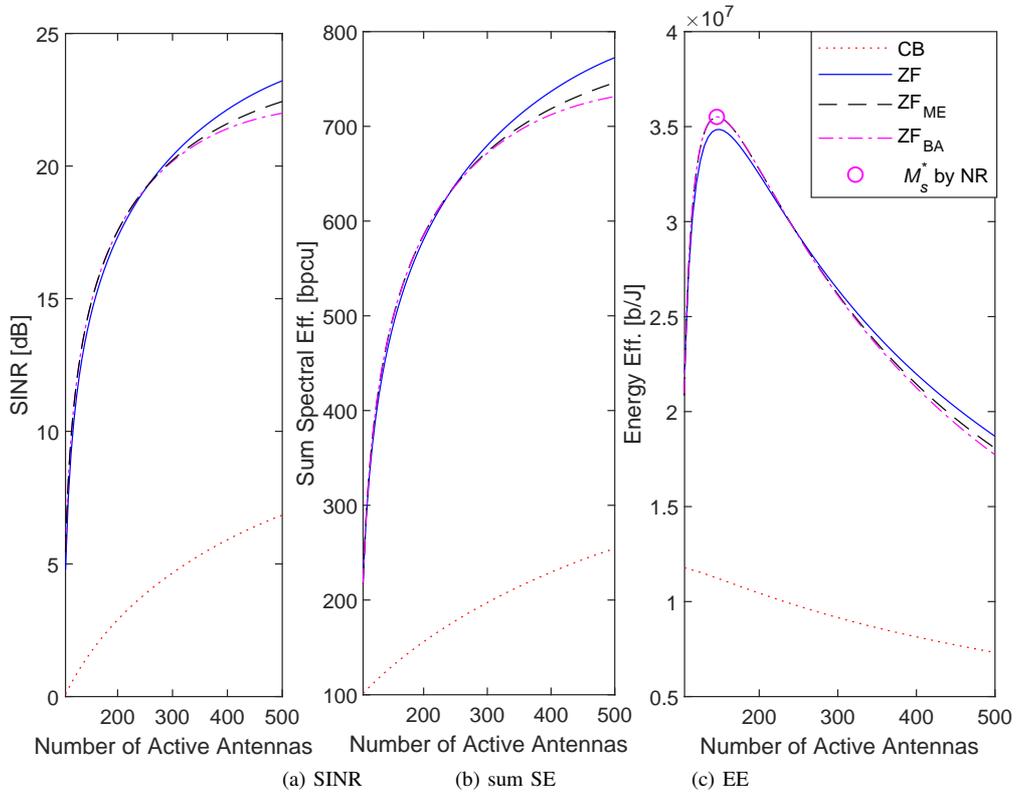} \\
\footnotesize (a) SINR \qquad \qquad (b) sum SE \qquad\qquad\qquad (c) EE
\vspace{-2mm}
\caption{HRNP-AS  scheme under ZF and CB precoders: (a) SINR, (b) sum SE, and (c) EE as a function of $M_s$ for $M = 500$ antennas and $K = 100$. }
\label{fig:SE_EE_selant}
\end{figure}

Next, in order to obtain the performance results of GA, LS, and PSO-based {AS} schemes, we have set the maximum number of iterations $N^{{\rm max}}_{it} = 60$ for such schemes, and analysed their convergence for $K=100$ users, as depicted in Fig. \ref{fig:Convergence}.a. One can see from the Figure that the LS-AS convergence presents a non-decreasing behavior, since when a new solution is not found in certain iteration, the algorithm interrupts its search, and does not spend more processing power. On the other hand, for GA and PSO-based AS for XL-MIMO systems, if the algorithms do not find new solutions and keep searching during additional iterations, the EE of that solution decreases due to the progressive processing power consumed in the subsequent iterations. Therefore, it is not efficient to predefine the number of iterations for these two schemes in the XL-MIMO antenna selection problem, since in this optimization problem it would be very difficult do adjust the number of iterations in such a way to obtain a suitable EE solution for the algorithms. To circumvent while taking advantage of this feature, we implement an {\it early-interruption} criterion, in which if the GA or the PSO-based AS schemes do not find a new solution within 5 iterations, the search is interrupted, obtaining the convergences depicted in Figure \ref{fig:Convergence}.b. Besides, for the GA-based {AS} scheme, we have considered a population size of $M/2$, of which $10\%$ are selected as parents at each iteration, and a mutation probability of $2\%$. For the PSO-based one, we have considered a swarm of $M/5$ particles, and an inertia weight, cognitive factor and social factor of 0.5.

\begin{figure}[htbp]
\centering
\includegraphics[width=.85\textwidth]{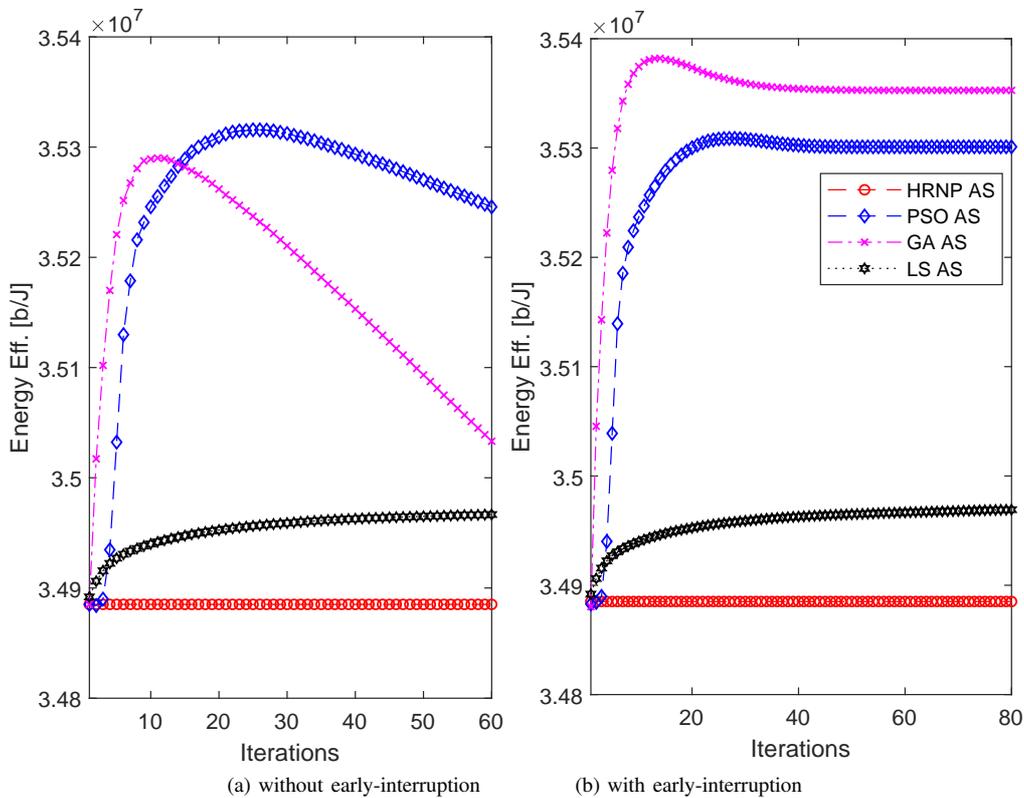} \\
\footnotesize (a) without early-interruption \qquad \qquad (b) with early-interruption 
\vspace{-2mm}
\caption{Convergence of the AS schemes: (a) without, and (b) with early-interruption stopping search criterion. $K=100$ and $M = 500$ antennas.}
\label{fig:Convergence}
\end{figure} 

Fig. \ref{fig:SE_EE_GAAS} depicts the SINR, sum SE, and EE as a function of $K$ for the HRNP, GA, LS, and PSO-based AS schemes employing ZF precoding, with $M = 500$ antennas at the XL-MIMO array. While the achieved sum SE performance is nearly the same for all investigated schemes, the graphs reveal that SINR and EE gains can be achieved in comparison with HRNP. The Figure also shows that, in terms of SINR and EE, the GA, LS, and PSO-based AS schemes achieve a similar performance, and their gains in comparison with HRNP AS are small, since the processing required for finding a suitable antennas subset in the XL-MIMO system increases the energy consumption; thus, the EE gains become marginal. Except for small number of users, the GA AS scheme achieves one of the best EEs in most part of the investigated scenario, although for high number of users, its performance becomes very similar to HRNP AS scheme. Besides, due to its simplicity and celerity to return the results, one can point out that the HRNP criterion coupled to the NR procedure for $M_s^*$ selection represents a very promising XL-MIMO AS scheme. 

\begin{figure}[!htbp]
\centering
\includegraphics[width=.85\textwidth]{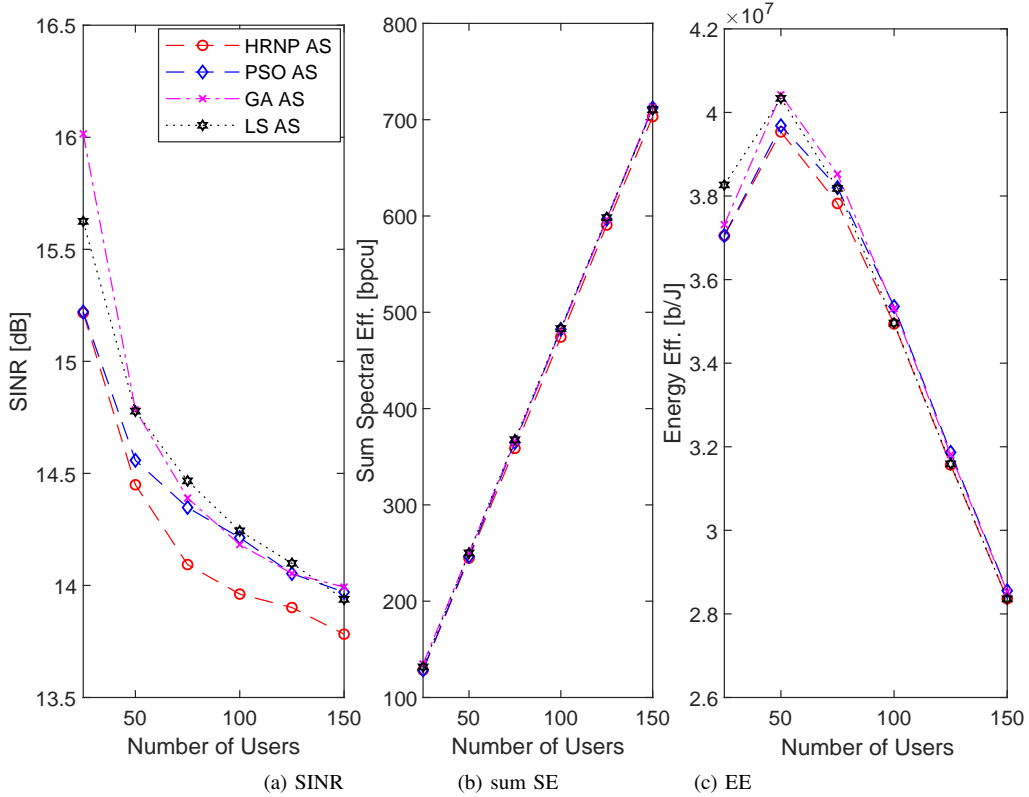}\\ \footnotesize (a) SINR \qquad \quad\qquad (b) sum SE \qquad\qquad\qquad (c) EE
\vspace{-2mm}
\caption{{AS schemes for ZF precoding: (a) SINR, (b) sum SE, and (c) EE as a function of $K$ for $M = 500$ antennas}. }
\label{fig:SE_EE_GAAS}
\end{figure} 

\subsection{Complexity of XL-MIMO AS Methods}\label{sec:compAS}
Fig. \ref{fig:Comp_Inc}.a depicts the average number of active antennas as a function of $K$ for the investigated AS methods. One can see that the $M_s^*$ value obtained by our proposed NR method usually matches the number of antennas selected by LS, PSO, and GA-based AS schemes, corroborating the tightness of the approximations made and the effectiveness of the method. The major advantage of our proposed NR method for obtaining $M_s^*$ is that it can be evaluated for any system configuration satisfying eq. \eqref{eq:cond_approx}. In our numerical simulations, the method has converged in at most 3 iterations from the starting point $M_{s,0} = 1.5 K$. Besides, the $M_s^*$ value is not dependent on the channel coefficients, but only on the system parameters, like number of users, transmit power, dimensions of XL-MIMO array and coverage area. Therefore, once found $M_s^*$, the NR method just has to be evaluated again when one of these parameters change. 
{The fixed complexity of evaluating $M_s^*$ under 3 NR iterations is about 380 \emph{flops}, which is negligible in comparison with that of selecting the antennas subset, eq. \eqref{eq:comphrnp}, \eqref{eq:compls}, \eqref{eq:compga}, and \eqref{eq:comppso}, besides of remaining valid for larger time periods.}

\begin{figure}[htbp]
\centering
\includegraphics[width=.85\textwidth]{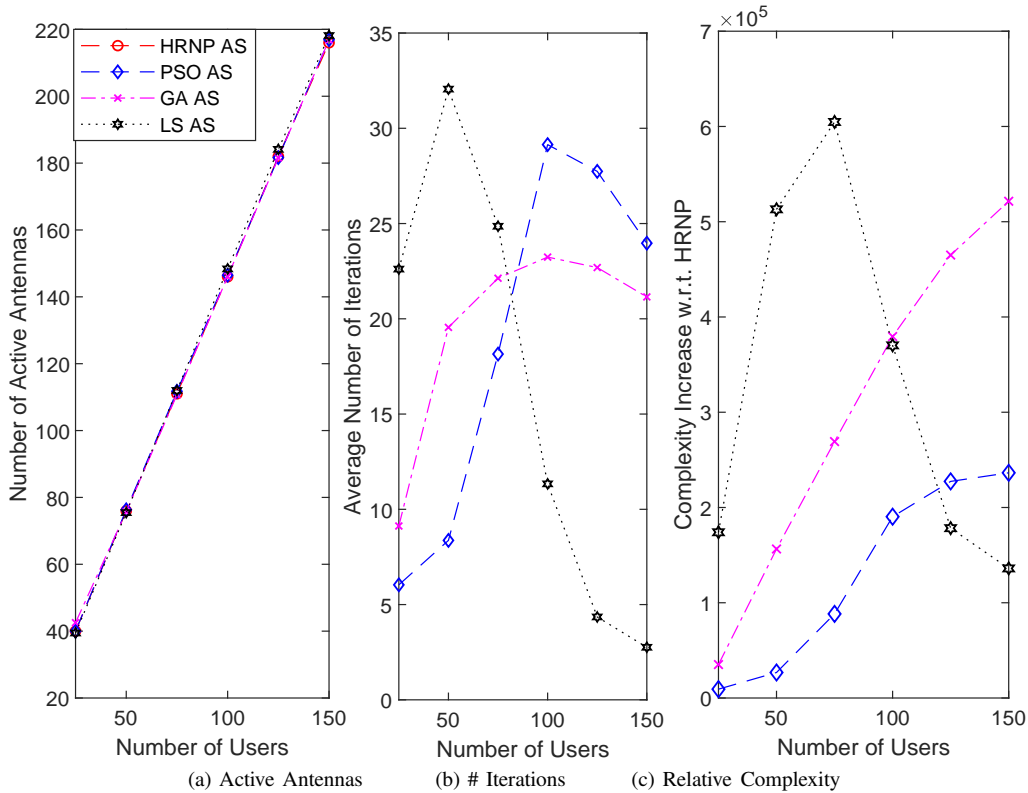} \\
\footnotesize (a) Active Antennas  \qquad\,\,\,\, (b)  \# Iterations \,\, \qquad (c) Relative Complexity
\vspace{-2mm}
\caption{(a) Average number of active antennas, (b) average number of iterations, and (c) complexity increase of the AS schemes w.r.t. HRNP AS approach as a function of $K$ for $M=500$ antennas.}
\label{fig:Comp_Inc}
\end{figure} 

Fig. \ref{fig:Comp_Inc}.b depicts the average number of iterations required by each investigated AS scheme, recalling that the number of iterations are not fixed, since the LS interrupts when a new solution is not find in an iteration, and GA and PSO implement the early-interruption criterion. Besides, due to the non-decreasing behavior of the LS convergence depicted in Fig. \ref{fig:Convergence}, the average number of iterations for this scheme in Fig. \ref{fig:Comp_Inc}.b does not correspond to the point in which the LS convergence curve becomes horizontal. Besides, the advantage of HRNP criterion in selecting  antennas within the XL-MIMO array can also be confirmed by the extra computational complexity required for the other analysed methods. Hence, considering the average number of iterations from Fig. \ref{fig:Comp_Inc}.b, the  {\it relative complexity increment} of the LS, GA and PSO AS schemes w.r.t. the HRNP AS method are depicted in Fig. \ref{fig:Comp_Inc}.c. The relative complexity increment metric is defined as:
$$
\Delta_{\mathcal{C}} =\frac{\mathcal{C}^{\textsc{ls, ga,pso}}_{\rm as}-\mathcal{C}^{\textsc{hrnp}}_{\rm as}}{\mathcal{C}^{\textsc{hrnp}}_{\rm as}}
$$
considering typical XL-MIMO network configurations for $K$ users and $M$ BS antennas. One can confirm the very large relative complexity increase of the AS methods for XL-MIMO, {\it i.e.}, this complexity increment is in the order of $10^5$, which make the benefits they would bring less significant in terms of energy efficiency. 

It is noteworthy that the computational complexity spent with the AS methods is included in the EE values, in terms of the processing power. In summary, the performance improvement of the AS scheme comes at the expense of high complexity, which results in marginal EE gains. On the other hand, the HRNP-AS procedure is able to achieve an improved EE of 34.85 Mbit/J for $K=100$ users, in comparison with 18.71 Mbit/J of selecting all antennas, {\it i.e.}, not applying any AS procedure, corresponding in a 86.3\% of EE increasing, as one can infer from Fig. \ref{fig:SE_EE_selant}.

{Elaborating further regarding the dependence of the optimal number of selected antennas $M_s^*$ on the system parameters, such as number of users, total transmit  power available,  dimensions  of XL-MIMO array,  and  coverage  area, one can argue that such system parameters vary quite slowly with respect to the data symbol period. Therefore, it could be possible to evaluate the proposed AS scheme, and turning-on the optimal number of RF chains $M_s^*$, which are then switched to the best antenna subset  according to our proposed HRNP criterion. Notice that only when the number of users changes significantly that it would be necessary to re-evaluate the \eqref{eq:Msf}-\eqref{eq:Msn}, and then turning-on or turning-off some RF chains. Besides, the proposed method for finding the optimal number of selected antennas can provide very useful information for XL-MIMO system designers.}

\section{Conclusion}\label{sec:concl}
In this paper, we have investigated the XL-MIMO systems subject to channel non-stationarities. First, we have revisited the performance expressions from \cite{Heath19}, and proposed to incorporate the power constraint at the SINR expressions of CB and ZF to arrive at more lean and comprehensive results. Then, based on such obtained expressions, we have proposed four XL-MIMO AS schemes aiming at maximizing the EE based on the following criteria: HRNP, LS, GA, and PSO. Some simplifying assumptions allowed us to derive closed-form EE expressions, based on which we proposed a NR iterative method to obtain the optimal number of active antennas. Numerical results have shown that GA usually achieves one of the best EEs, although the gains were marginal in comparison with HRNP, since the processing required for achieving a suitable antennas subset increases the consumed energy, limiting the achieved EE gains. Thus, due to its simplicity and celerity in returning results, the proposed HRNP-AS scheme, with the NR method providing the optimal subarray size value $M_s^*$, can be seen as a very promising solution for AS XL-MIMO systems, achieving an EE gain of 86.3\% in comparison with selecting all antennas strategy.

\end{document}